\documentclass[aps,prl,twocolumn,nopacs,superscriptaddress]{revtex4}
\usepackage[utf8]{inputenc}

\usepackage{graphicx}  
\usepackage{dcolumn}   
\usepackage{bm}        
\usepackage{amssymb}   
\usepackage{amsmath}
\usepackage{units}
\usepackage[dvipsnames]{xcolor}
\usepackage{xfrac}

\usepackage{sidecap}
\sidecaptionvpos{figure}{m}  

\usepackage[type1]{libertine}                                        
\usepackage{textcomp}
\usepackage[scaled=.85]{beramono}
\usepackage[libertine,cmintegrals,cmbraces,vvarbb,slantedGreek]{newtxmath}
\usepackage[scr=boondoxo]{mathalfa}
\usepackage{bm}
\usepackage[lf]{carlito}

\usepackage{braket}
\usepackage{bm}

\makeatletter
\DeclareRobustCommand{\cev}[1]{%
  \mathpalette\do@cev{#1}%
}
\newcommand{\do@cev}[2]{%
  \fix@cev{#1}{+}%
  \reflectbox{$\m@th#1\vec{\reflectbox{$\fix@cev{#1}{-}\m@th#1#2\fix@cev{#1}{+}$}}$}%
  \fix@cev{#1}{-}%
}
\newcommand{\fix@cev}[2]{%
  \ifx#1\displaystyle
    \mkern#23mu
  \else
    \ifx#1\textstyle
      \mkern#23mu
    \else
      \ifx#1\scriptstyle
        \mkern#22mu
      \else
        \mkern#22mu
      \fi
    \fi
  \fi
}
\makeatother
\newcommand{\eqdef}{\overset{\mathrm{def}}{=\joinrel=}}

\newcommand{\spinhalf}{spin-\sfrac{1}{2}}

\renewcommand{\Im}{\text{Im}}

\begin{document}

\title{Spin-dependent tunneling between individual superconducting bound states}

\author{Haonan Huang}
\affiliation{Max-Planck-Institut f\"ur Festk\"orperforschung, Heisenbergstraße 1,
70569 Stuttgart, Germany}
\author{Jacob Senkpiel}
\affiliation{Max-Planck-Institut f\"ur Festk\"orperforschung, Heisenbergstraße 1,
70569 Stuttgart, Germany}
\author{Ciprian Padurariu}
\affiliation{Institut für Komplexe Quantensysteme and IQST, Universität Ulm, Albert-Einstein-Allee 11, 89069 Ulm, Germany}
\author{Robert Drost}
\affiliation{Max-Planck-Institut f\"ur Festk\"orperforschung, Heisenbergstraße 1,
70569 Stuttgart, Germany}
\author{Alberto Villas}
\affiliation{Departamento de F\'{\i}sica Te\'orica de la Materia Condensada and
Condensed Matter Physics Center (IFIMAC), Universidad Aut\'onoma de Madrid, 28049 Madrid, Spain}
\author{Raffael L. Klees}
\affiliation{Fachbereich Physik, Universität Konstanz, 78457 Konstanz, Germany}
\author{Alfredo Levy Yeyati}
\affiliation{Departamento de F\'{\i}sica Te\'orica de la Materia Condensada and
Condensed Matter Physics Center (IFIMAC), Universidad Aut\'onoma de Madrid, 28049 Madrid, Spain}
\author{Juan Carlos Cuevas}
\affiliation{Departamento de F\'{\i}sica Te\'orica de la Materia Condensada and
Condensed Matter Physics Center (IFIMAC), Universidad Aut\'onoma de Madrid, 28049 Madrid, Spain}
\author{Bj\"orn Kubala}
\affiliation{Institut für Komplexe Quantensysteme and IQST, Universität Ulm, Albert-Einstein-Allee 11, 89069 Ulm, Germany}
\author{Joachim Ankerhold}
\affiliation{Institut für Komplexe Quantensysteme and IQST, Universität Ulm, Albert-Einstein-Allee 11, 89069 Ulm, Germany}
\author{Klaus Kern}
\affiliation{Max-Planck-Institut f\"ur Festk\"orperforschung, Heisenbergstraße 1,
70569 Stuttgart, Germany}
\affiliation{Institut de Physique, Ecole Polytechnique Fédérale de Lausanne, 1015 Lausanne, Switzerland}
\author{Christian R. Ast}
\email[Corresponding author; electronic address:\ ]{c.ast@fkf.mpg.de}
\affiliation{Max-Planck-Institut f\"ur Festk\"orperforschung, Heisenbergstraße 1,
70569 Stuttgart, Germany}

\date{\today}

\begin{abstract}

Magnetic impurities on superconductors induce discrete bound levels inside the superconducting gap, known as Yu-Shiba-Rusinov (YSR) states. YSR levels are fully spin-polarized such that the tunneling between YSR states depends on their relative spin orientation. Here, we use scanning tunneling spectroscopy to resolve the spin dynamics in the tunneling process between two YSR states by experimentally extracting the angle between the spins. To this end, we exploit the ratio of thermally activated and direct spectral features in the measurement to directly extract the relative spin orientation between the two YSR states. We find freely rotating spins down to 7\,mK, indicating a purely paramagnetic nature of the impurities. Such a non-collinear spin alignment is essential not only for producing Majorana bound states but also as an outlook manipulating and moving the Majorana state onto the tip.

\end{abstract}

\maketitle

\vspace{0.5cm}

Magnetic impurities on superconductors give rise to a wealth of phenomena due to the subtle interplay between pairing processes of conduction electrons and their exchange interaction with the impurity \cite{Pan2000,Nadj-Perge2014,Menard2017,Kezilebieke2019,Odobesko2020,Kezilebieke2020}. As a consequence, single spin-nondegenerate bound states inside the superconducting gap may emerge \cite{Yu1965,Shiba1968,Rusinov1969,Yazdani1997,Ji2008,Franke2011,Hatter2017,Malavolti2018,Senkpiel2019,Huang2019magnetic}. A fundamental understanding of the role of spin in the coupling between such states is highly relevant \cite{Saldana2018}, for example, in the context of creating Majorana bound states at the ends of a chain of Yu-Shiba-Rusinov (YSR) states \cite{kim2018toward,kamlapure2018engineering}. Aside from the conventional quasiparticle hopping from one site to the next, pair creation from the superconducting condensate can also couple two neighboring sites in a chain of YSR states. To reach the topological regime in such a chain, the coexistence of both types of coupling processes between neighboring sites is indispensable \cite{Kitaev2001,Alicea2012}. The spin provides a vital component selecting between the two coupling processes. Conventional hopping is favored when the spins of two sites are aligned, while pair creation is favored when neighboring spins are anti-aligned.

Past experiments observed end states on magnetic atom chains on superconductors using scanning tunneling microscopy \cite{Nadj-Perge2014,kamlapure2018engineering,kim2018toward}, but the two tunneling processes between neighboring sites in the chain could not be individually accessed. To separate the two tunneling processes, control over the energy detuning between adjacent sites is needed. For this purpose, instead of using the tip as an \textit{external} sensor \cite{kamlapure2018engineering,Kezilebieke2018}, measuring a YSR state with a functionalized YSR tip turns the tunnel junction into an \textit{internal} probe of the coupling between two YSR sites.

\begin{figure}
    \centering
    \includegraphics[width=\columnwidth]{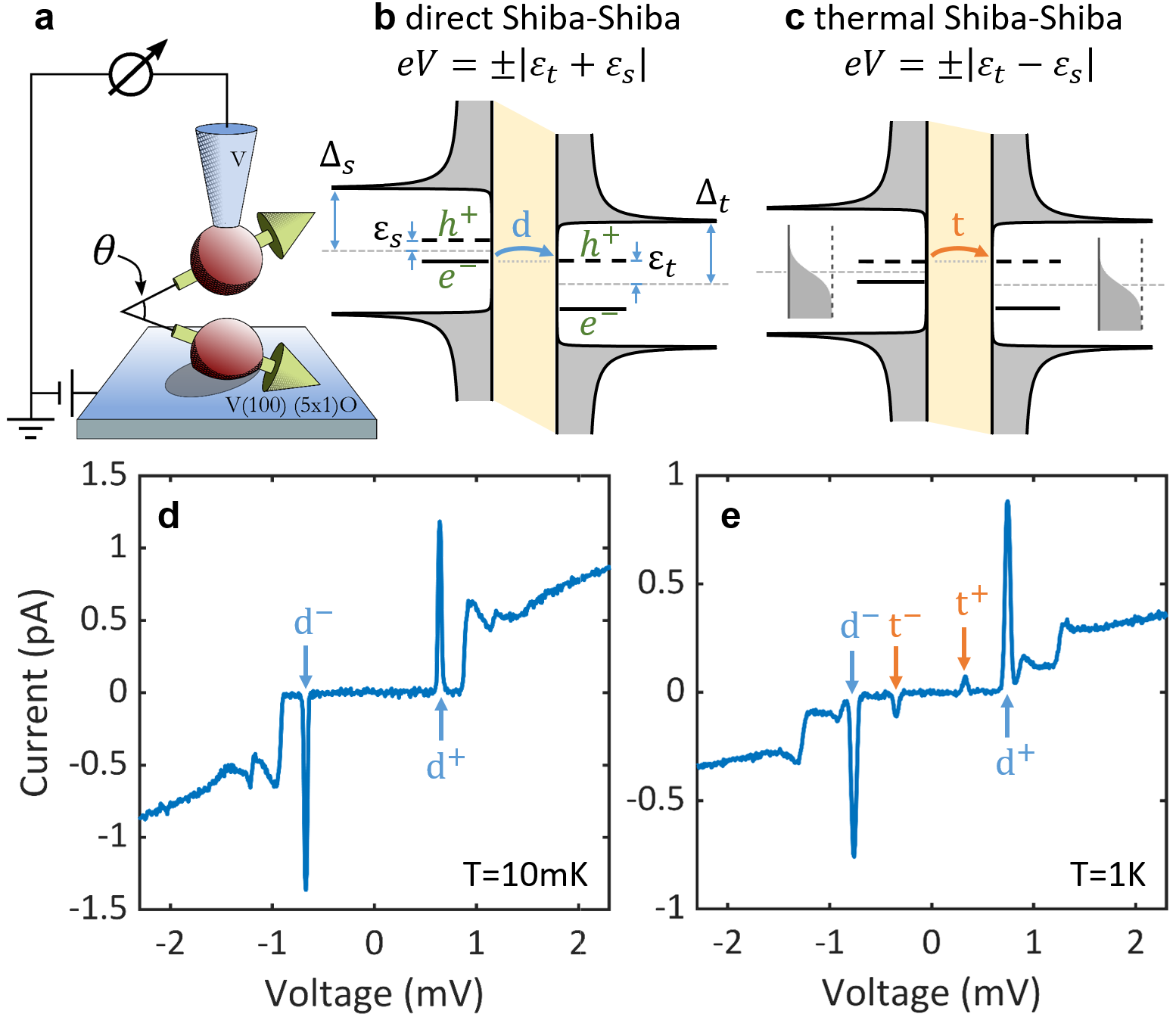}
    \caption{\textbf{Direct Shiba-Shiba tunneling and thermal Shiba-Shiba tunneling.} (a) Schematics of the experimental setup realizing tunneling between Yu-Shiba-Rusinov (YSR) states using a scanning tunneling microscope, with $\theta$ denoting the angle between the two spins. b,c) Tunneling processes of direct (b) and thermal (c) Shiba-Shiba tunneling. In the direct process, the $e^-$ part and the $h^+$ part of the two YSR states are aligned by the bias voltage. In the thermal process, the two $e^-$ parts or two $h^+$ parts are aligned. $\Delta_{s,t}$ are the superconducting gap parameters of the sample and tip respectively, with both around 760$\,\mu$eV for vanadium. (d),(e) Typical $I(V)$ spectra of Shiba-Shiba tunneling at 10\,mK (d) and 1\,K (e). Blue arrows ($d^{\pm}$) indicate direct Shiba-Shiba peaks, and orange arrows ($t^{\pm}$) indicate thermal Shiba-Shiba peaks.}
    \label{fig_1}
\end{figure}

Here, we use a scanning tunneling microscope (STM) \cite{Assig2013} to couple a YSR state on the vanadium tip and an intrinsic YSR state on the V(100) surface, where we independently resolve direct and thermally activated Shiba-Shiba tunneling (Fig.\ \ref{fig_1}(a-c)). We first show experimentally from the spin blockade of a characteristic Andreev reflection that an individual state is spin-nondegenerate. Measuring the two tunnel processes between two spin-nondegenerate superconducting bound states, we find that both direct and thermal processes are present for all measured impurities. Their coexistence excludes the possibility that the impurity spins are collinear, fulfilling the requirement for use in topological chains. The measured relative strength of the two tunnel processes enables us to extract the angle $\theta$  of the relative spin orientation. Tracing the relative spin orientation from 1\,K to 7\,mK, we find $\theta = 90^{\circ}$ in all measurements indicating freely rotating spins of the YSR states in tip and sample with no detectable anisotropy. At the end, we draw an analogy between Shiba-Shiba tunneling and the coupling of neighboring states in a YSR chain.

When both tip and sample exhibit a YSR state, the tunneling between these discrete levels can be observed in the current-voltage $I(V)$ characteristic as isolated spectral features \cite{Huang2020tunneling}. At 10\,mK, we only observe current peaks at bias voltages indicated by blue arrows labeled $d^{\pm}$ in Fig.\ \ref{fig_1}(d), which we refer to as \textit{direct Shiba-Shiba tunneling}. This process is found at a bias voltage $V$ of the sum of two YSR energies $eV=\pm |\epsilon_t+\epsilon_s|$ ($e$ is the electron charge, $\epsilon_{t,s}$ are YSR energies for the tip and sample, respectively). The corresponding process is plotted schematically in Fig.\ \ref{fig_1}(b).

At higher temperature (1\,K), however, the YSR state may get thermally excited, such that thermally activated tunneling becomes possible. Figure \ref{fig_1}(c) shows schematically how an excitation tunnels from one side to the other. Additional current peaks appear at the bias voltage $eV=\pm |\epsilon_t-\epsilon_s|$ (see Fig. \ref{fig_1}(e) indicated by orange arrows and labeled $t^{\pm}$), which we call \textit{thermal Shiba-Shiba tunneling}. The intensity of the thermal Shiba-Shiba peaks is smaller compared to the direct Shiba-Shiba peak due to the small thermal Boltzmann factor at 1\,K. Due to their appearance at distinct bias voltages, the direct and thermal Shiba-Shiba processes can be separately addressed in a single measured $I(V)$ spectrum. While the general dynamics of the tunneling processes of Shiba-Shiba tunneling has been discussed recently \cite{Huang2020tunneling}, the role of the spin in the transport process is more subtle, but plays a defining part in the coupling between YSR states.

\begin{figure}
    \centering
    \includegraphics[width=\columnwidth]{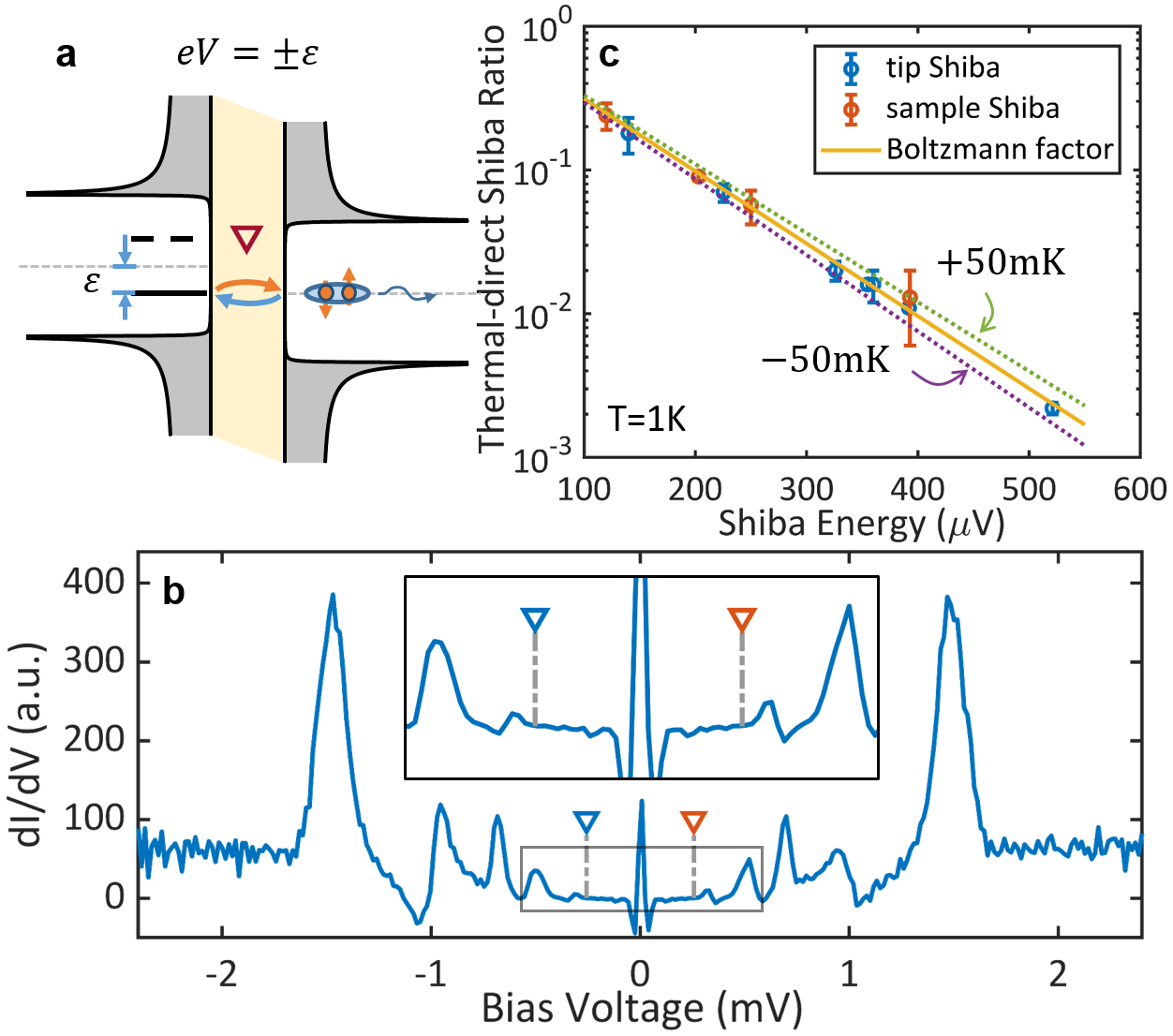}
    \caption{\textbf{Properties of the YSR states.} (a) A family of multiple Andreev Reflection processes (MARs) connecting the levels of the same YSR state (the schematic here show only the lowest order process). Due to full spin polarization of the YSR state, this family of MARs is spin forbidden. (b) A typical dI/dV spectrum involving one YSR state tunneling into a clean superconductor at high conductance (here $\sim 0.16 G_0$, where $G_0=2e^2/h$ is the conductance quantum) to reveal MARs. Detailed peak assignment can be found in the supplementary information \cite{supinf}. The expected position for the lowest order MARs in (a) is labeled with symbol $\triangledown$, where no peak is seen confirming the spin blockade. (c) Statistics of the thermal-direct ratio of the conventional YSR-BCS peak (in the low conductance regime) of various YSR impurities (tip or sample) with different YSR energy measured at 1K. All data points fall on top of the prediction of the Boltzmann factor at 1K within a narrow window ($\pm 50\,\mathrm{mK}$). }
    \label{fig_2}
\end{figure}

\begin{figure}
    \centering
    \includegraphics[width=\columnwidth]{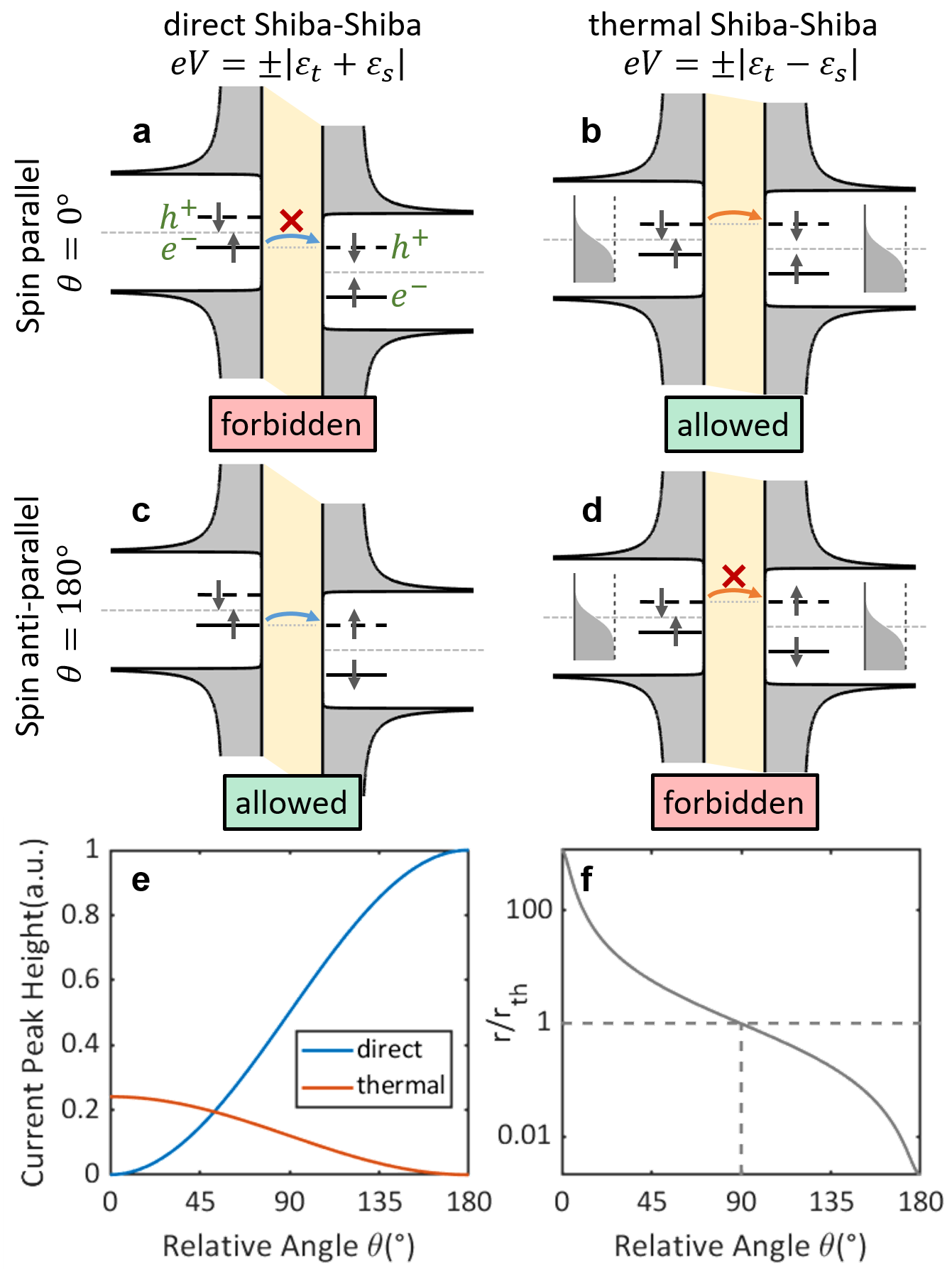}
    \caption{\textbf{Theory showing thermal-direct Shiba-Shiba ratio as a measure of the relative spin orientation $\theta$.} (a,b) The tip and sample YSR states have parallel spins ($\theta=0^\circ$). Note that the $e^-$ part and $h^+$ part of the same YSR state always have opposite spins \cite{Balatsky2006,Cornils2017,schneider_atomic-scale_2020}. Here, the direct Shiba-Shiba process is forbidden while thermal Shiba-Shiba tunneling is allowed due to spin conservation during tunneling. (c,d) The tip and sample YSR states have opposite spins ($\theta=180^{\circ}$). Here, the direct Shiba-Shiba process is allowed while thermal Shiba-Shiba tunneling is forbidden. (e) The continuous dependency of the direct and thermal Shiba-Shiba peak intensity on the relative spin angle. (f) The universal dependency of $r/r_\textrm{th}$ on the relative spin angle.}
    \label{fig_3}
\end{figure}

Conventional models describe YSR states as spin-nondegenerate states, i.e.,\ 100\% spin polarized \cite{Shiba1968,Balatsky2006}, but not necessarily with a preferred axis. By contrast, a dominant Kondo coupling with a Kondo temperature $T_\text{K}$ larger than the superconducting gap $\Delta$ ($T_\text{K}\gg\Delta)$ may lead to a spin-degenerate bound state. Experimentally, it should thus be verified that an alleged YSR resonance actually is spin polarized \cite{Cornils2017,schneider_atomic-scale_2020}. Without resorting to magnetic fields, this can be checked by looking at the presence or absence of the single Andreev reflection process involving the same bound state \cite{Villas2020} as shown in Fig.\ \ref{fig_2}(a). The absence of such a spectral feature at a bias voltage of $eV=\pm\epsilon$ indicates that the bound state is a spin-nondegenerate YSR state \cite{Villas2020}. A high conductance spectrum of a single YSR state in the tunnel junction featuring multiple Andreev reflection peaks is shown in Fig.\ \ref{fig_2}(b). At the expected bias voltage, the respective single Andreev process is absent in the measurement, i.e.,\ there is no spectral feature at the voltage indicated by the symbol $\triangledown$, which confirms the spin-nondegeneracy of the YSR state (for the assignment of all observed peaks and other details see the supplementary information \cite{supinf}).

In order to understand the influence of spin on the Shiba-Shiba peak intensities, we first consider the limiting case of the YSR states being aligned either parallel ($\theta=0^\circ$) or anti-parallel ($\theta=180^\circ$) as shown in Figs.\ \ref{fig_3}(a)-(d). If the spins of the corresponding parts of the YSR states are parallel, direct Shiba-Shiba tunneling is forbidden (Fig.\ \ref{fig_3}(a)). This is due to spin conservation in the tunneling process and because the electron ($e^-$) and hole ($h^+$) parts, which participate in the tunneling process, have opposite spins. The thermal Shiba-Shiba process, however, is allowed, because no spin blockade exists (Fig.\ \ref{fig_3}(b)) for tunneling between two hole parts due to parallel spins.

If the spins of the YSR states are anti-parallel, the situation is reversed. Using similar arguments as above, direct Shiba-Shiba tunneling is now allowed (Fig.\ \ref{fig_3}(c)), while thermal Shiba-Shiba tunneling is forbidden (Fig.\ \ref{fig_3}(d)) (for a detailed discussion see supplementary information \cite{supinf,Shiba1968,Balatsky2006,Cornils2017}).

Observing both direct and thermal Shiba-Shiba peaks in Fig.\ \ref{fig_1}(e) directly leads to the conclusion that the spins in the measurement can neither be completely parallel nor completely anti-parallel. Using the standard expression for the tunneling current explicitly including the spin \cite{supinf}
\begin{equation}
    I(V) = \sum\limits_{\sigma} \left(I_{\sigma\sigma}(V)\cos^2\frac{\theta}{2} + I_{\sigma\bar{\sigma}}(V)\sin^2\frac{\theta}{2}\right),
    \label{eq:ivspin}
\end{equation}
where $\sigma=\uparrow,\downarrow$ and $\bar{\sigma}=\downarrow,\uparrow$, the $I(V)$ spectra can be simulated expanding the above discussion to arbitrary angles between the spins of the two YSR states ($0^{\circ}<\theta<180^{\circ}$) (for details see supplementary information \cite{supinf}). The result is shown in Fig.\ \ref{fig_3}(e). The Shiba-Shiba peak intensities evolve smoothly between the extreme cases of parallel and anti-parallel orientations, confirming the limiting cases discussed before. In the measurement, we find two peaks for each direct and thermal process (one at positive and one at negative bias voltages). To account for both peaks in the analysis, we average between them using the geometric mean which eliminates the electron-hole asymmetry and the absolute intensity of the peaks, so that the resulting ratio $r$ only depends on the angle $\theta$ between the spins of the two YSR states as well as the Boltzmann factors of the two YSR states
\begin{equation}
    r\eqdef\sqrt{\frac{p_{t^+}p_{t^-}}{p_{d^+}p_{d^-}}}= \cot^2\left(\frac{\theta}{2}\right)\left|e^{-\epsilon_\text{s}/k_\mathrm{B}T}-e^{-\epsilon_\text{t}/k_\mathrm{B}T}\right|,
    \label{eq:r}
\end{equation}
where $T$ is the temperature, $k_\text{B}$ is the Boltzmann constant, $p_i$ with $i=d^{\pm},t^{\pm}$ is the intensity of the corresponding Shiba-Shiba peak. Further, we define a parameter $r_\text{th}$ as the difference of the two Boltzmann factors, which are experimentally accessible, such that we find
\begin{equation}
    \frac{r}{r_\mathrm{th}}= \cot^2\left(\frac{\theta}{2}\right).
    \label{eq:ratio}
\end{equation}
Note that here it is possible to use either current peak area or peak height, because the width of direct and thermal Shiba-Shiba is the same (for details see supplementary information \cite{supinf}).

The predicted evolution of $r/r_\text{th}$ is plotted in Fig.\ \ref{fig_3}(f), which maps a unique angle for every ratio. If $r/r_\mathrm{th} \rightarrow \infty$, the spins are parallel; if $r/r_\mathrm{th} \rightarrow 0$, spins are anti-prallel. Equation \eqref{eq:ratio} presents a universal curve that is independent of the details of each YSR state. Because the Shiba-Shiba peaks are typically well separated in the measured spectrum, both $r$ and $r_\text{th}$ in Eq.\ \eqref{eq:ratio} can be precisely derived from directly measurable quantities in a single measurement, such that the angle $\theta$ can be experimentally determined.
It is this ability to independently resolve the relative strength of two different Shiba-Shiba tunneling processes in the $I(V)$ spectrum, which enables an all-electrical measurement of the relative spin-orientation in our experiment -- a striking difference to conventional spin-valve experiments, where a polarizing magnetic field is used to rotate (or flip) the relative orientation.

To extract the Boltzmann factors, we consider experimentally the thermal to direct ratio of the YSR state tunneling into the continuum \cite{Ruby2015a}. A statistics of such measurements at 1K is plotted in Fig.\ \ref{fig_2}(c) (for details see the supplementary information \cite{supinf}). All Boltzmann ratios for both tip and sample YSR states lie on a line corresponding to $1\,\text{K}\pm0.05\,$K. This agrees with our thermometer reading and also confirms that we have only one pair of YSR states inside the gap \cite{Ruby2015a}.

\begin{figure}
    \centering
    \includegraphics[width=\columnwidth]{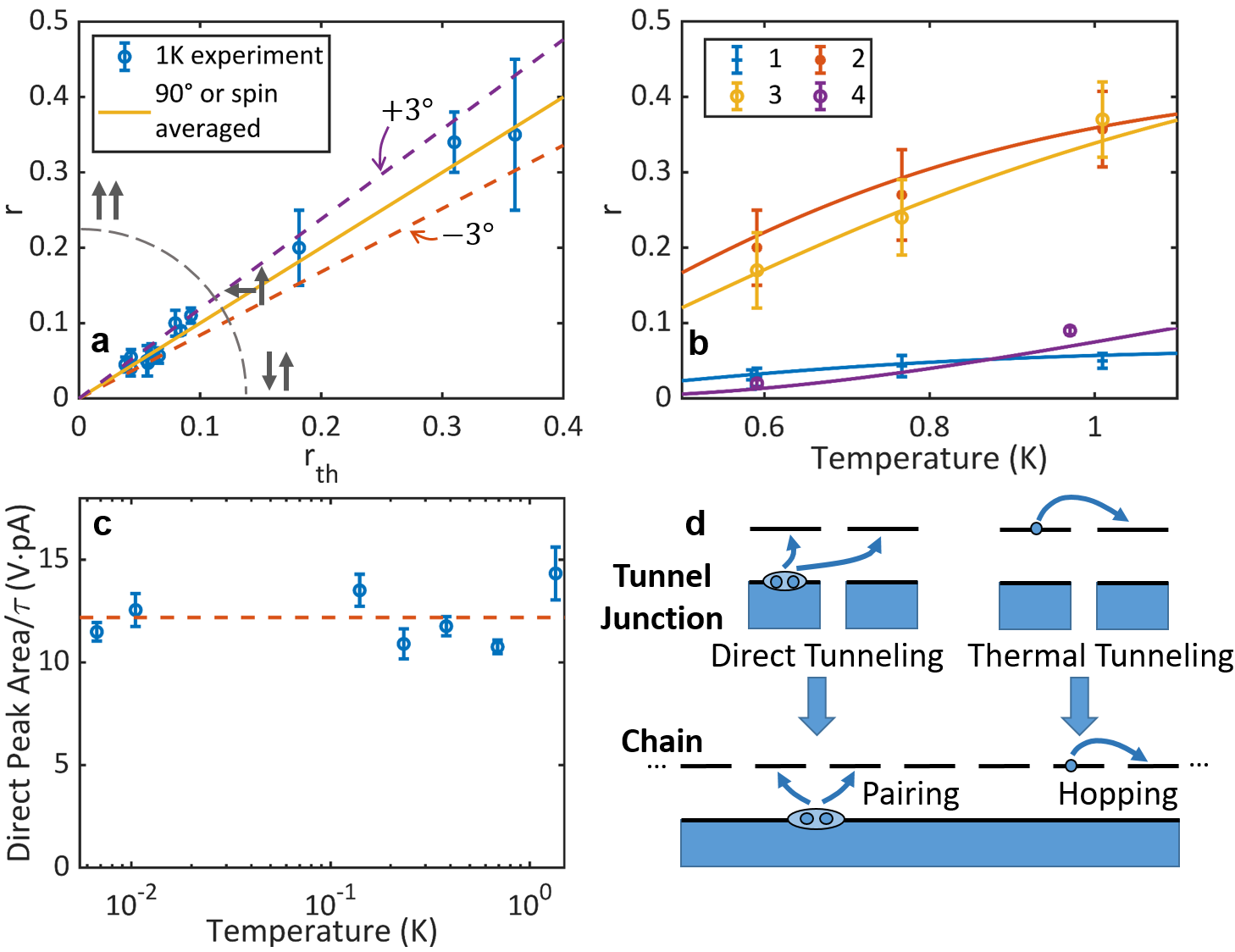}
    \caption{\textbf{Temperature dependency measurements.} (a) Statistics of the thermal-direct Shiba-Shiba ratio $r$ (of peak current) measured at 1K, plotted against the difference of the Boltzmann factor $r_\textrm{th}$ of tip and sample Shiba state. All data points lie on the identity line within a small error bar, indicating a spin averaging effect. (b) The temperature dependence of thermal-direct Shiba-Shiba ratio $r$ of selected examples with different YSR energies (data points) agreeing with the prediction from the difference of the Boltzmann factor $r_\textrm{th}$ (curves of corresponding color). (c) The temperature dependence of the direct Shiba-Shiba current peak area normalized by the transmission $\tau=G_N/G_0$, which remains nearly constant, with no indication of spin freezing down to 7\,mK. (d) The direct and thermal tunneling between two YSR states constitute the fundamental interaction mechanism of the formation of a chain of magnetic impurities on a superconductor. The co-existence of both processes is crucial for the chain to host topologically non-trivial end states.}
    \label{fig_4a}
\end{figure}

We have collected the experimental parameters $r$ and $r_\text{th}$ from a number of different Shiba-Shiba systems in the linear tunneling regime using intrinsic defects on the V(100) surface and a V tip at 1\,K. Note that in the tunneling regime the two-state system is minimally disturbed as higher order tunneling processes are suppressed. In Fig.\ \ref{fig_4a}(a), we plot the thermal-direct Shiba-Shiba ratio $r$ against $r_\mathrm{th}$. Interestingly, for all Shiba-Shiba systems, the data points lie on the identity line ($r=r_\mathrm{th}$), within a small error interval. According to Eq.\ \eqref{eq:ratio}, this means that the the relative angle between the two YSR spins is $\theta=90^{\circ}\pm3^{\circ}$. Considering that \textit{every} Shiba-Shiba system in Fig.\ \ref{fig_4a}(a) shows the same angle $\theta$, it is unlikely that these angles are time-independent. Instead, we are measuring the average angle of a time dependent spin orientation. The tunneling time is short compared to any motion of the spins, such that every tunneling event will observe a static snapshot of the system, so that the measured current is a time average over many independent tunneling events. Therefore, we propose that the spins are freely rotating, giving an average effect $\braket{r}=r_\mathrm{th}$, where $\braket{}$ indicates averaging over an angle isotropically distributed over the unit sphere (for more details, see the supplementary information \cite{supinf}). Note that although the spin is freely rotating, the electron and hole part of the same YSR state has opposite spin direction at any time.

A temperature of 1\,K may be still too high for a magnetic system to develop an easy axis \cite{Loth2012}, which could explain the observed isotropy. We, therefore, trace the parameters $r$ and $r_\text{th}$ as a function of temperature down to 0.6\,K. Figure \ref{fig_4a}(b) shows the thermal-direct Shiba-Shiba peak ratio $r$ for four different Shiba-Shiba systems between 0.6\,K and 1\,K. Upon cooling, the data points for the ratio $r$ nicely follow the calculation from Eq.\ \eqref{eq:r} (solid lines) assuming an angle $\theta = 90^{\circ}$. This indicates that $r=r_\mathrm{th}$ with a still freely rotating spin scenario down to 0.6\,K.

If we cool further down below 0.6\,K, the ratio becomes more difficult to access because of the exponential suppression of thermal Shiba-Shiba peak. However, recalling Fig. \ref{fig_3}(e), we note that the direct Shiba-Shiba peak intensity by itself critically depends on the relative spin angle $\theta$. To eliminate the conductance dependency as well as temperature effects on the YSR lifetime and environmental broadening \cite{Huang2020tunneling,Ast2016,Devoret1990,Averin1990,Ingold1994}, we use the peak area normalized by the conductance (see supplementary information for details \cite{supinf}). Figure \ref{fig_4a}(c) shows a typical data set of the normalized direct Shiba-Shiba peak area from 1\,K down to 7\,mK, showing no significant change over two decades of temperature range. These measurements are consistent with a relative spin orientation rotating isotropically down to 7\,mK without any anisotropy energy ($k_\mathrm{B}T<0.6\,\mu\text{eV}$). We find that for YSR states, the relative orientation of the spins determines whether direct or thermal tunneling is possible (cf.\ Fig.\ \ref{fig_3}). This is different from the spin valve effect in normal conducting tunnel junctions, where an antialignment suppresses the tunneling current, but does not favor a different transport channel.

It is tempting to draw an analogy between the two coupled YSR states in our experiment and a link in a chain of YSR states \cite{Nadj-Perge2014}. The two types of coupling processes, direct and thermal Shiba-Shiba tunneling, are counterparts of the pair creation and hopping terms in the Kitaev Hamiltonian (see Fig. \ref{fig_4a}(d)) \cite{Kitaev2001,Huang2020tunneling,supinf}. In this analogy, Majorana end states require the coexistence of both coupling processes, which is guaranteed by the non-collinearity of the neighboring impurity spins. We surmise that the relative spin orientation between YSR states is equally important in the chain as in the tunnel junction. One difference, however, is that in our experiment the two impurities are placed on different superconducting substrates relatively far away compared to the atomic chain. This rules out magnetic interactions via the substrate, which have been suggested to induce helical order in Kitaev chains \cite{pientka2013,Pientka2014,Hoffman2016,Andolina2017,Klinovaja2013,vazifeh2013self,Braunecker2013}. Nevertheless, as an outlook, our results suggest an intriguing experiment to probe the edge of a magnetic impurity chain by a YSR tip. The tip YSR state will couple to the state at the edge of the chain by direct and thermal Shiba-Shiba tunneling and extend the chain by one site. The Majorana end state will move from the substrate onto the tip.

In summary, we propose the ratio between the thermal and direct Shiba-Shiba tunneling peaks as a direct and unambiguous all-electrical probe for the relative spin orientation between YSR states. We conclude that for the Shiba-Shiba systems consisting of intrinsic YSR impurities on V(100) and YSR states on the vanadium tip, the YSR states derive from freely rotating quantum \spinhalf\ impurities from 1\,K down to 7\,mK. The spin plays a defining role in the transport between spin-nondegenerate superconducting bound states selecting between hopping and pairing mechanisms (i.e.,\ direct and thermal Shiba-Shiba tunneling). Additionally, the analogy between the Shiba-Shiba junction and a chain link in a YSR chain attributes a critical role to the spin for producing Majorana end states. Further, a YSR tip in close proximity to a Majorana end state may induce a topological phase transition allowing the Majorana end state to be manipulated by the tip. Going beyond the immediate consequences, a Shiba-Shiba junction may also provide the basis for a minimal source to inject triplet Cooper pairs into a superconductor \cite{eschrig_theory_2003}.

\section*{Acknowledgments}
We would like to acknowledge fruitful discussions with Gianluca Rastelli and Wolfgang Belzig. This work was funded in part by the ERC Consolidator Grant AbsoluteSpin (Grant No.\ 681164) and by the Center for Integrated Quantum Science and Technology (IQ$^\textrm{\small ST}$). J.A. acknowledges funding from the DFG under grant number AN336/11-1. A.V., A.L.Y. and J.C.C. acknowledge funding from the Spanish MINECO (Grant No. FIS2017-84057-P and FIS2017-84860-R), from the ``Mar\'{\i}a de Maeztu'' Programme for Units of Excellence in R\&D (MDM-2014-0377). R.L.K. acknowledges support by the DFG through SFB 767 and Grant No.\ RA 2810/1.

\clearpage
\widetext
\begin{center}
\textbf{\large Supplementary Material for \\ Spin-dependent tunneling between individual superconducting bound states}
\end{center}

\setcounter{figure}{0}
\setcounter{table}{0}
\setcounter{equation}{0}
\renewcommand{\thefigure}{S\arabic{figure}}
\renewcommand{\thetable}{S\Roman{table}}
\renewcommand{\theequation}{S\arabic{equation}}

\vspace{0.5cm}

\section{Materials and methods}

\subsection{Tip and sample preparation}
The sample was a V(100) single crystal with $>99.99\%$ purity, and was prepared by standard UHV metal preparation procedure of multiple cycles of argon ion sputtering around $10^{-6}\,$mbar argon pressure with about 1\,keV acceleration energy and annealing at around $700^\circ$C. The sample was heated up and cooled down slowly, around $1^\circ$C/s, to reduce strain in the crystal. The tip was made from a polycrystalline vanadium wire of 99.8\% purity, which was cut in air and prepared in ultrahigh vacuum by Argon sputtering. Subsequent field emission on V(100) surface as well as standard tip shaping techniques were used to obtain a tip exhibiting clean bulk gap as well as good imaging capabilities.

\subsection{YSR tip preparation}
By controlled indentation, we can induce a YSR state to the apex of the vanadium tip, shown in detail in Ref.\ \cite{si_Huang2020tunneling}. We perform multiple cycles of dipping the tip into the sample to exchange material between tip and surface. In this way, YSR states with desired energy, intensity and asymmetry can be produced on the tip. The tip YSR state likely comes from a combination of simple elements such as oxygen and carbon that are picked up from the surface, but the exact mechanism requires further investigation experimentally and theoretically \cite{si_Huang2020tunneling}.

\subsection{Reaching different temperature with the mK-STM}
The mK-STM is capable of reaching temperatures between 7\,mK and 1\,K. It has two continuous operation modes: one at 1\,K where nearly all $^3$He--$^4$He mixture is taken out from the dilution refrigerator, another at 10\,mK where nearly all mixture is condensed in the mixing chamber of the dilution refrigerator. To reach around 0.6\,K and 0.8\,K, we can start from the 1\,K mode and put a little bit mixture inside the cycle to increase the cooling power. To reach a temperature from 10\,mK to around 500\,mK continuously, we can start from 10\,mK and apply some resistive heating in the mixing chamber.

To reach below 10\,mK, we can continue pumping the mixture but stop the mixture input into the mixing chamber to reduce heat load on the mixing chamber. This is a single shot mode because once the ${}^3 \mathrm{He}$ is pumped empty, the mixing chamber will warm up. Before that, the temperature can be lowered to between 6\,mK and 7\,mK.

\section{Experimental details}

\subsection{V(100) surface and intrinsic YSR defects}
Vanadium is a type-II conventional Bardeen-Cooper-Schrieffer (BCS) superconductor, with a transition temperature $T_\text{C}=5.4\,$K and a gap parameter $\Delta\cong 760\,\upmu$eV \cite{si_Sekula1972,si_Zasadzinski1982}. Our preparation of tip and sample repeatably yields the measured gap parameter close to the bulk value for both tip and sample. In the temperature range between 7\,mK  and 1\,K, both tip and sample are well superconducting. The V(100) surface typically shows a $(5\times 1)$ oxygen reconstruction due to the diffusion of bulk oxygen impurities to the surface through sputtering-annealing cycles during sample preparation \cite{si_Jensen1982,si_Koller2001,si_Dulot2001}. The existence of such an oxygen layer does not have any effect on the superconductivity measured on the surface \cite{si_Jack2016}.

There are various types of impurities on the surface, with oxygen vacancies and carbon probably the most abundant \cite{si_Huang2020tunneling}, which show no magnetic signature for the most cases. There is a sparse distribution of intrinsic magnetic impurities (on the order of 0.04\% of a monolayer) that show YSR states inside the superconducting gap on the surface, which is observed consistently throughout different samples or preparation cycles. The abundance of such impurities, however, does depend on the annealing temperature. The exact origin of such impurities requires further research, but they probably come from some uncommon arrangements of simple elements such as oxygen and carbon \cite{si_Huang2020tunneling}.

\subsection{Spin-forbidden YSR-MARs}

\begin{SCfigure*}
    \centering
    \includegraphics[width=0.7\textwidth]{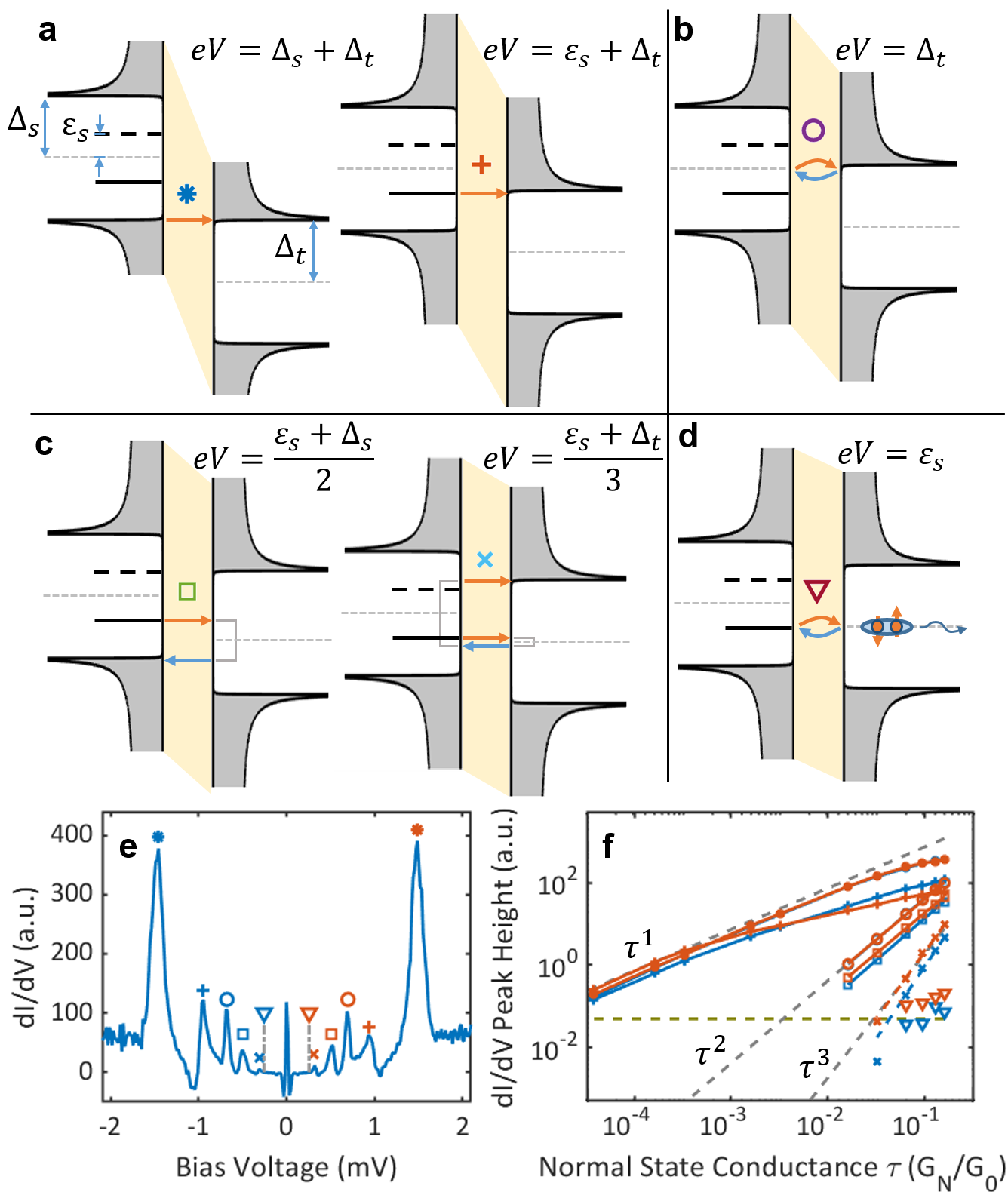}
    \caption{\textbf{Multiple Andreev Reflections (MARs) with YSR states.} (a) Direct quasiparticle tunneling processes (no Andreev reflection). Left ($\ast$): tunneling between continua. Right ($+$): tunneling from YSR state to the continuum. (b) MARs without involving the YSR state. Here in the diagram ($\circ$), the second order MAR is shown (one reflection). (c) MARs connecting YSR levels to coherence peaks. Left ($\square$): second order process (one Andreev reflection). Right ($\times$): third order process (two Andreev reflections). (d) MARs connecting levels of the same YSR state. Here in the diagram ($\triangledown$): second order process (one reflection). This family of MARs is strictly forbidden due to the full spin polarization of the YSR state. (e) (main text Fig. 2(b)) $dI/dV$ spectrum of a YSR state on the sample surface at the normal state conductance of $0.16G_0$, with all peaks labeled with the same marks in (a)-(d) indicating the origin of each peak. The reverse triangles $\triangledown$ denote the expected energy position for the process depicted in (d), which is forbidden resulting in the absence of the peak. (f) The conductance dependence of the corresponding peaks from (e), showing different order $\tau^n$. The horizontal dahsed line is the noise limit. If there were peaks following the process in (d), the reverse triangles would follow a $\tau^2$ dependence, which is not the observation here.}
    \label{fig_si_1}
\end{SCfigure*}

The spin polarization of YSR states has been shown previously by spin polarized STM \cite{si_Cornils2017}, but here we use the signature in multiple Andreev Reflections (MARs) to demonstrate the full spin polarization of YSR states without needing a spin polarized probe (nor magnetic field). MARs are higher order tunneling processes that arise when one or both electrodes are superconducting. MARs are responsible for many subgap features in the tunneling spectra between two superconductors. Depending on the number of quasiparticles being transferred across the tunnel junction, the intensity of the associated spectral feature scales with the conductance $G_N$ according to a power law $\propto G_N^i, i=\{2,3,4...\}$, where $i$ is the order of MAR. The order of MAR is defined here as the number of quasiparticles crossing the interface and, therefore, the lowest order MAR is second order (the simple Andreev reflection) with one reflection. Therefore, from the peak position as well as the conductance dependence of a certain peak in the $dI/dV$ spectrum, we can unambiguously attribute the relevant MAR and its order.

Due to the superlinear conductance dependence, MARs only become visible at relatively high conductance values. In the main part of this paper (Shiba-Shiba tunneling), we worked at very low conductance so that no MAR features are involved, simplifying the analysis to extract the relative spin angle. Here, however, we are going to show in the relatively high conductance regime that MARs can be utilized to demonstrate the full spin polarization of YSR states. The experimental data shown in Figs.\ \ref{fig_si_1}(e,f) comes from the tunneling between an intrinsic sample YSR state on the V(100) surface and a clean superconducting vanadium tip, where the gap parameters are $\Delta_t=700 \mathrm{\mu eV}, \Delta_s=760 \mathrm{\mu eV}$ for the tip and sample respectively and the YSR energy being $\epsilon = 260 \mathrm{\mu eV}$ extracted from the spectrum. Figure \ref{fig_si_1} (e) (i.e., main text Fig.\ 2(b), reproduced here for easier discussion and comparison) is measured at the normal state conductance of around $0.16G_0$. All labels ($\ast$, $+$, $\circ$, etc.) across Figs. \ref{fig_si_1}(a)-(f) are consistent so that for a certain process, its transition diagram (Figs. \ref{fig_si_1}(a)-(d)), spectral peak (Fig. \ref{fig_si_1}(e)) and the conductance dependency (Fig. \ref{fig_si_1}(f)) can be directly associated.

In the case of one YSR state on the sample, multiple MAR processes can occur (Fig.\ \ref{fig_si_1}) which have been discussed in detail in theory \cite{si_Villas2020}. First, there are direct quasiparticle processes (Fig.\ \ref{fig_si_1}(a)) responsible for the $dI/dV$ peaks at the gap edge (labeled with $\ast$) as well as the conventional YSR-BCS tunneling peak (labeled with $+$). These processes scale linearly with $G_N$ at low conductance but enter a sublinear regime at high conductance (Fig. \ref{fig_si_1}(f)), which has been discussed before in the context of multiple tunneling events and lifetime effect \cite{si_Ruby2015a,si_Huang2020tunneling}. The second family is conventional MARs without involving the YSR states (Fig.\ \ref{fig_si_1}(b)), and the peaks labeled with $\circ$ in Fig.\ \ref{fig_si_1}(e) is attributed to be a lowest order MAR (second order) that scales quadratically with the conductance (Fig. \ref{fig_si_1}(f)). The third family comprises MARs connecting the YSR level with the clean superconductor coherence peaks (Fig.\ \ref{fig_si_1}(c)), where the second order (labeled with $\square$) as well as third order (labeled with $\times$) processes are observed in Figs. \ref{fig_si_1}(e),(f) featuring the expected conductance dependence. These MARs are usually quite strong because of the spectroscopic significance of YSR levels compare to the coherence peaks. This family of MARs does not feature any spin dependence, because it tunnels into the continuum, which is spin degenerate.

The fourth family is exotic because it starts from and ends in a YSR state (Fig.\ \ref{fig_si_1}(d), labeled with $\triangledown$). Imagine a spin up electron tunnels to the right, and in order to generate a Cooper pair on the right hand side, it needs to take one spin down electron from the left to pair with. This process is not allowed because the YSR state is fully polarized and the electron part has only one spin species. Apart from this intuitive description, it has been shown in theory that this family of MARs is strictly forbidden due to the spin polarization of the YSR state \cite{si_Villas2020}. If the YSR state were a spin degenerate level or not fully spin polarized, we would expect to see the peaks originating from this family in the spectrum. These features would be quite prominent due to YSR state being nearly a singularity in the density of states. As a side note, if the quasiparticle background of the superconductor is non-negligible, there could be a peak at the expected energy position, but will scale linearly with conductance, because it is direct quasiparticle tunneling from the YSR level to the continuous quasiparticle background of the other electrode rather than MARs \cite{si_randeria_scanning_2016,si_Villas2020}.

In the experiment, we do observe the above spin filtering. We have seen all lowest order MARs (second order) as well as one third-order MAR in the other families, but there is no peak at the expected position for the lowest order process of the fourth MARs family (Figs. \ref{fig_si_1}(d),(e) $\triangledown$). Plotting the $dI/dV$ signal at the expected position against conductance in Fig.\ \ref{fig_si_1}(f) ($\triangledown$) does not show the expected $\propto G_N^2$ dependence. Consequently, without applying a magnetic field or using a spin-polarized probe, we demonstrate that the YSR states are fully spin polarized by showing that this "self Shiba-Shiba tunneling" (fourth family of MARs) is strictly forbidden.

\subsection{Shiba-Shiba tunneling}

Although Shiba-Shiba tunneling results in multiple peaks in the narrow superconducting gap even in the case of single YSR state on each electrode, the spectral features are generally well separated allowing for unambiguous attribution. Since $0<\epsilon_{t,s}<\Delta_{t,s}$, we have $|\epsilon_t-\epsilon_s| < |\epsilon_t+\epsilon_s| < |\epsilon_{t,s}+\Delta_{s,t}|$, where $\Delta_{s,t}$ are the superconducting gap parameter of the sample and the tip, respectively. $eV=|\epsilon_{t,s}+\Delta_{s,t}|$ is the lowest bias voltage possible to drive a direct tunneling process into the quasiparticle continuum outside the superconducting gap, which corresponds to the conventional tunneling from a YSR state into the continuum outside the gap of the other electrode. Consequently, the thermal Shiba-Shiba peak ($eV = \pm|\epsilon_t-\epsilon_s|$) is closest to zero voltage, separated from the direct Shiba-Shiba peak($eV = \pm|\epsilon_t+\epsilon_s|$), while all other features are more outside. This becomes more complicated if the temperature is higher or the conductance is higher, where more peaks are expected inside the gap due to thermal excitations or higher order processes.

\subsection{The relative spin orientation in relation to the quantum phase transition}

The spins of the YSR levels depend not only on the impurity spin, but also on which side of the quantum phase transition (QPT) the system is. When a YSR state moves across a QPT, e.g.,\ when the spin-dependent scattering strength crosses a threshold, the $e^--h^+$ asymmetry will be exchanged, accompanied with reversed polarization of the levels above and below the Fermi energy \cite{si_Shiba1968, si_Balatsky2006,si_Cornils2017}. Consequently, the spin-parallel case in the discussion in the main text corresponds to two possibilities, either the impurity spins are parallel and the YSR states are on the same side of the QPT, or impurities have opposite spins and the YSR states are on the opposing side of the QPT. The spin-anti-parallel case would correspond to the opposing combinations, i.e.,\ parallel impurity spins but at opposing sides of the QPT, or anti-parallel impurity spins at the same side of the QPT.

\subsection{Experimentally extracting the ratio $r$ and $r_\mathrm{th}$}

\begin{SCfigure*}
    \centering
    \includegraphics[width=0.72\textwidth]{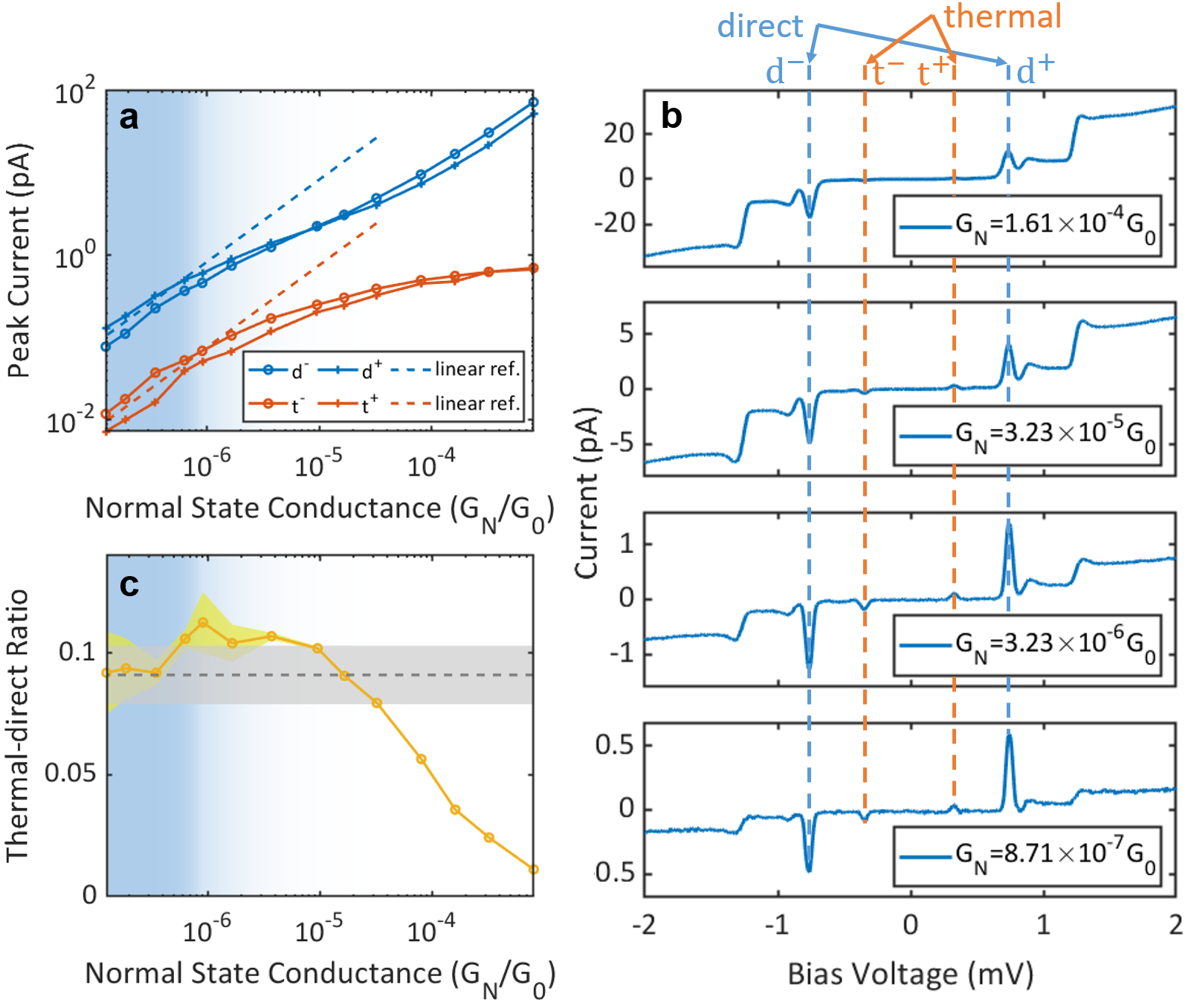}
    \caption{\textbf{Conductance dependence measurements of the Shiba-Shiba tunneling.} (a,b) Conductance dependency of thermal and direct Shiba-Shiba tunneling. (a) The evolution of the Shiba-Shiba current peak height with respect to normal state conductance, showing the linear regime in the low conductance limit as blue shaded area. (b) Selected $I(V)$ spectra at different conductance, with thermal and direct Shiba-Shiba peaks labeled respectively. (c) Conductance dependency of the thermal-direct Shiba-Shiba ratio, with the low conductance value indicated by the gray dashed line being the value $r$ used in this paper. The vertical shadings are the estimated uncertainty.}
    \label{fig_si_2}
\end{SCfigure*}

The experimental extraction of the thermal-direct Shiba-Shiba ratio $r=\sqrt{\frac{p_{t^+}p_{t^-}}{p_{d^+}p_{d^-}}}$ should be done under appropriate tunneling conditions, because the theoretical model shown in the main text and in more detail later here in the supplementary material is valid only in the low conductance limit (linear regime \cite{si_Huang2020tunneling}). At higher conductance, resonant processes as well as multiple Andreev reflections make the interpretation more complicated \cite{si_Villas2020}. Therefore, in order to extract the Shiba-Shiba peak intensities correctly, the system has to be in the linear regime, where the Shiba-Shiba peak intensities depend linearly on the tunneling conductance and the YSR states have enough time to relax into the ground state before the next tunneling event. As the intrinsic lifetime of the YSR states varies from system to system, it is not \textit{a priori} obvious below which conductance threshold the system behaves linearly, so a conductance dependency measurement is necessary \cite{si_Huang2020tunneling}. To illustrate this, we have plotted the Shiba-Shiba peak currents for a typical YSR tip and YSR sample system in Fig.\ \ref{fig_si_2}(a) as a function of the normal state conductance. The normal state conductance is extracted from the conductance measurement at 4\,mV, which is much larger than the superconducting gap. The measurement extends several orders of magnitude of the conductance to resolve the different regimes. The peak current values have been extracted as indicated in some selected spectra in Fig.\ \ref{fig_si_2}(b). Only in the blue shaded area in Fig.\ \ref{fig_si_2}(a), the system behaves linearly. Beyond this conductance threshold, higher-order effects start to emerge forcing the system into a sublinear conductance dependence. As a result, to measure $r$ correctly, one needs a conductance dependence experiment and take $r$ in the linear regime (shaded area in Fig.\ \ref{fig_si_2}(c)).

The second parameter $r_\mathrm{th}$ can be extracted from the experiment in one of two ways. One way is to directly obtain the temperature $T$ from the thermometer reading and the YSR energies $\epsilon_{s,t}$ from the position of the peaks in the spectra. The other method to determine $r_\mathrm{th}$ experimentally is independent of the first method. It relies on the ratio between the peak intensities of the direct and thermal transitions in the single impurity YSR tunneling into the continuum. Here, the ratio between the thermal peak (at $eV=\pm (\Delta_{s,t}-\epsilon_{t,s})$) and the direct peak (at $eV=\pm (\Delta_{s,t}+\epsilon_{t,s})$) in the differential conductance is simply the Boltzmann factor $e^{-\epsilon_{t,s}/k_\mathrm{B}T}$  (see theory part below and \cite{si_Ruby2015a}). Fig. 2(c) of the main text shows the result of the ratio for various tip and sample YSR states separately at 1\,K as a function of the respective YSR state energy using the latter procedure, showing good consistency. Consequently, we can characterize the Boltzmann factor of the sample YSR state by measuring the spectrum using a clean superconducting tip, and measure that of the tip YSR state by moving the YSR tip to a clean spot on the sample without YSR states and measure the spectrum. After these measurements, $r_\mathrm{th}$ in the Shiba-Shiba tunneling is then simply the difference between the two measured Boltzmann factors as in Eq.\ \eqref{eq_ratio_4by4}.

\subsection{Error estimation of the thermal-direct Shiba-Shiba peak ratio}
The uncertainty $\sigma_r$ of the ratio $r=\sqrt{\frac{p_{t^+}p_{t^-}}{p_{d^+}p_{d^-}}}$ shown as shading in Fig. \ref{fig_si_2}(c) comes from three factors. First, the measured Shiba-Shiba peak current $p_i$ (especially the thermal Shiba-Shiba peak $p_t$) in the linear regime can be as low as 10\,fA so that the noise from the I-V converter becomes significant in the determination of the current peak height. Second, the linear regime may locate at quite small conductance making its determination difficult, which is especially challenging at very low temperature. Third, such a set of measurements takes time due to long sweep average as well as multiple spectra for conductance dependency, and relocating the YSR defect after drift may contribute to additional error.

The first error contribution can be quantitatively characterized, and the error estimate of each single point in Fig. \ref{fig_si_2}(c) (yellow shading) considers this factor only, with the formula $\sigma_r=\sqrt{\Sigma_{i=\{t^+,t^-,d^+,d^-\}}(\frac{\partial r}{\partial p_i})^2 \sigma^2_{p_i}} = \frac{r\sigma_p}{2}\sqrt{\Sigma_{i=\{t^+,t^-,d^+,d^-\}}(\frac{1}{p_i})^2}$, where $\sigma_{p}$ is the uncertainty of the current measurement extracted from the spectrum. We assume that the current noise is independent of the bias voltage because at such low current, the noise from the I-V converter dominates which is independent of the bias.

The second and third error contribution can only be estimated roughly, and the total error estimate of $r$ (gray shading in Fig.\ \ref{fig_si_2}(c) and subsequently error bars in Fig.\ 2(c) and Figs. 4(a),(b) of the main text) also try to include these effects. Notice that the ratio is not prone to the third error because the thermal YSR and direct YSR state usually have similar spatial extension, but the absolute intensity depends critically on the spatial position, which may explain the deviation from constant of the temperature dependency of direct Shiba-Shiba peak area in Fig.\ 4(c) in the main text.

\subsection{Shiba-Shiba tunneling below 0.6\,K}
The ratio becomes difficult to access below 0.6\,K because of the exponential suppression of thermal Shiba-Shiba peak. In addition, the YSR lifetime is enhanced when cooling down due to the suppression of thermal relaxation channels, resulting in the reduction of the linear regime to even lower conductance (cf.\ Fig. \ref{fig_si_2}(a)), making both Shiba-Shiba processes harder to resolve. Therefore, the direct Shiba-Shiba peak intensity is used to trace the relative spin angle $\theta$.
Notice, however, unlike the ratio where both peak area and peak height can be used, only current peak area can be used here to eliminate the temperature effect on the YSR lifetime and environmental broadening \cite{si_Huang2020tunneling,si_Ast2016,si_Devoret1990,si_Averin1990,si_Ingold1994}. The reason is that environmental broadening as well as external noise preserves the peak area.

\section{Green's function theory of Shiba-Shiba tunneling}

To simulate the tunneling spectra, we use the non-equilibrium Green's function theory formalism. First, we work in the $2 \times 2$ Nambu space to demonstrate some basic properties, then we introduce the $4 \times 4$ spin-Nambu space which is essential for arbitrary spin orientations.

\subsection{General formalism in the 2$\times$2 Nambu space}

We describe a general superconductor-superconductor tunnel contact with the Hamiltonian \cite{si_Cuevas1996}
\begin{equation}
    \hat{H}(\tau) = \hat{H}_L + \hat{H}_R + \sum_{\sigma}\Big( te^{i\phi(\tau)/2}c^\dagger_{L\sigma}c_{R\sigma }+t^{*}e^{-i\phi(\tau)/2} c^\dagger_{R\sigma}c_{L\sigma } \Big),
\end{equation}
where $\hat{H}_{L,R}$ are the unperturbed Hamiltonians of the two electrodes, $t$ is the coupling parameter, $\phi(\tau)$ is the time-dependent superconducting phase with $\phi(\tau)=\phi_0+\frac{2\omega_0\tau}{\hbar}, \omega_0 = eV$, and $V$ is the bias voltage applied across the junction.

The general expression of current between two electrodes can be written in the time domain
\begin{equation}
    I(\tau)=\frac{e}{\hbar}\mathrm{Tr}[\tau_3(\hat{t}(\tau)G_{RL}^{+-}(\tau,\tau)-G_{LR}^{+-}(\tau,\tau)\hat{t}^{\dagger}(\tau))],
    \label{eq_current_ori}
\end{equation}
in which $\tau_3$ is the Pauli z matrix in Nambu space, $\mathrm{Tr}$ is the trace over Nambu degrees of freedom, and $\hat{t}(\tau)$ is the time-dependent hopping $\hat{t}(\tau)=\begin{pmatrix}
te^{i\phi(\tau)/2} & 0 \\
0 & -t^{*}e^{-i\phi(\tau)/2}
\end{pmatrix}
$.

We introduce the lesser Green's function $G^{+-}(\tau,\tau)$ (in some literature written as $G^{<}$):
\begin{equation}
G_{ij}^{+-}(\tau,\tau')=i\begin{pmatrix}
\langle c_{j\uparrow}^\dagger(\tau')c_{i\uparrow}(\tau)\rangle &
\langle c_{j\downarrow}(\tau')c_{i\uparrow}(\tau)\rangle \\
\langle c_{j\uparrow}^\dagger(\tau')c_{i\downarrow}^\dagger(\tau)\rangle &
\langle c_{j\downarrow}(\tau')c_{i\downarrow}^\dagger(\tau)\rangle
\end{pmatrix}.
\end{equation}
All matrices are in $2\times 2$ Nambu space, and with some algebra we can write the expression of current in terms of individual matrix elements of the Green's functions
\begin{equation}
    I(\tau)=\frac{e}{\hbar}(te^{i\omega_0 \tau/\hbar}G_{RL,11}^{+-}-t^*e^{-i\omega_0 \tau/\hbar}G_{LR,11}^{+-}
    +t^*e^{-i\omega_0 \tau/\hbar}G_{RL,22}^{+-}-te^{i\omega_0 \tau/\hbar}G_{LR,22}^{+-}).
\label{eq1}
\end{equation}

The Green's functions involved have the following relations \cite{si_Cuevas1996}

\begin{eqnarray}
    G_{RL}^{+-}(\tau,\tau)=\int d\tau_1[G_{RR}^r(\tau,\tau_1)\hat{t}^\dagger(\tau_1)g_{L}^{+-}(\tau_1-\tau)+G_{RR}^{+-}(\tau,\tau_1)\hat{t}^\dagger(\tau_1)g_{L}^a(\tau_1-\tau)],\label{eq2}\\
    G_{LR}^{+-}(\tau,\tau)=\int d\tau_1[g_{L}^{+-}(\tau-\tau_1)\hat{t}(\tau_1)G_{RR}^{a}(\tau_1,\tau)+g_{L}^{r}(\tau-\tau_1)\hat{t}(\tau_1)G_{RR}^{+-}(\tau_1,\tau)],\label{eq3}
\end{eqnarray}
in which $g$ stands for the unperturbed Green's functions and $G$ for the dressed Green's functions carrying all influence of the other electrode.

Generally, with a few more equations, the current can be solved numerically, which yields the linear to sublinear transition of Shiba-BCS tunneling and Shiba-Shiba tunneling, as well as multiple Andreev reflections (MARs) \cite{si_Cuevas1996,si_Villas2020}. To simplify and to get analytical results, however, we consider here the low conductance limit, which corresponds to the linear regime, where the current is proportional to $|t|^2$.

\subsection{The tunneling current in the linear regime}

In this limit, all dressed Green's functions of one electrode can be replaced with unperturbed Green's functions:

\begin{equation}
G_{RR}^{+-}=g_{R}^{+-},\; G_{RR}^{r,a}=g_{R}^{r,a}.
\end{equation}

Consequently, Eqs. \eqref{eq2} and \eqref{eq3} become:
\begin{eqnarray}
    G_{RL}^{+-}(\tau,\tau)&=&\int d\tau_1[g_{R}^r(\tau-\tau_1)t^\dagger(\tau_1)g_{L}^{+-}(\tau_1-\tau)+g_{R}^{+-}(\tau-\tau_1)t^\dagger(\tau_1)g_{L}^a(\tau_1-\tau)],\label{eq7}\\
    G_{LR}^{+-}(\tau,\tau)&=&\int d\tau_1[g_{L}^{+-}(\tau-\tau_1)t(\tau_1)g_{R}^{a}(\tau_1-\tau)+g_{L}^{r}(\tau-\tau_1)t(\tau_1)g_{R}^{+-}(\tau_1-\tau)].\label{eq8}
\end{eqnarray}

To obtain the DC current, we perform a Fourier transform, and with the following relations between Green's functions
\begin{eqnarray}
G^a-G^r&=&G^{+-}-G^{-+},\\
g^{+-}&=&(g^a-g^r)f(E),\label{eq4}\\
g^{-+}&=&-(g^a-g^r)(1-f(E)),\label{eq5}
\end{eqnarray}
where $f(E)=\frac{1}{e^{E/{k_\mathrm{B}T}}+1}$ is the Fermi function and $G^{-+}$ is the greater Green's function ($G^{>}$ in some literature), we arrive at the following expression for the DC current:
\begin{align}
I_0=\frac{e|t|^2}{h}\int_{-\infty}^{+\infty} &d\omega [g_{L,11}^{+-}(\omega)g_{R,11}^{-+}(\omega+\omega_0)-g_{L,22}^{+-}(\omega)g_{R,22}^{-+}(\omega-\omega_0)\nonumber\\
&-g_{R,11}^{+-}(\omega)g_{L,11}^{-+}(\omega-\omega_0)+g_{R,22}^{+-}(\omega)g_{L,22}^{-+}(\omega+\omega_0)].
\label{eq6}
\end{align}

We can further simplify the result by denoting the density of states $\rho_{ii}(\omega)=\mp\frac{1}{\pi}\Im{[g^{r,a}_{ii}(\omega)]},i\in \{1,2\}$,
\begin{equation}
I_0=\frac{4\pi^2 e|t|^2}{h}\int_{-\infty}^{+\infty}  \big(\rho_{L,11}(\omega)\rho_{R,11}(\omega+\omega_0)+\rho_{L,22}(\omega+\omega_0)\rho_{R,22}(\omega)\big)\big(f(\omega)-f(\omega+\omega_0)\big) d\omega,
\label{eq6.2}
\end{equation}
which is the convolution of the densities of states.

\subsection{Green's function of a YSR state}
The Hamiltonian of a YSR state in the classical Shiba model can be written as \cite{si_Shiba1968,si_Balatsky2006}
\begin{equation}
    H=\sum_{k\sigma}\xi_k c_{k\sigma}^\dagger c_{k\sigma} + \sum_{k}(\Delta c_{k\uparrow}^\dagger c_{-k\downarrow}^\dagger+h.c.)+\sum_{\sigma}(JS\sigma+U)c_{0\sigma}^\dagger c_{0\sigma},
    \label{eq_Hamil}
\end{equation}
where $J$ is the exchange coupling, $S$ is the spin and $U$ is potential scattering. The YSR Green's function in the $2\times 2$ Nambu space can be derived from the above Hamiltonian
\begin{equation}
g_\text{YSR}(\omega)=\frac{\pi \nu_0}{2\omega\alpha-(1-\alpha^2+\beta^2)\sqrt{\Delta^2-\omega^2}}\begin{pmatrix}
\omega+(\alpha+\beta)\sqrt{\Delta^2-\omega^2} & -\Delta \\
-\Delta & \omega+(\alpha-\beta)\sqrt{\Delta^2-\omega^2}
\end{pmatrix},\label{eq_YSR}
\end{equation}
with dimensionless parameters $\alpha=\pi\nu_0 JS$ and $\beta=\pi\nu_0 U$, and $\nu_0$ contains the normal-state density of state at the Fermi energy. When $\alpha=0, \beta=0$, it reduces to the usual BCS Green's function
\begin{equation}
g_\text{BCS}(\omega)=\frac{-\pi \nu_0}{\sqrt{\Delta^2-\omega^2}}\begin{pmatrix}
\omega & -\Delta \\
-\Delta & \omega
\end{pmatrix}.\label{eq_bcs}
\end{equation}
A substitution of $\omega \rightarrow \omega+i\eta$ results in the retarded Green's function $G^r$ and a substitution of $\omega \rightarrow \omega-i\eta$ results in the advanced Green's function $G^a$, where $\eta$ is a small real number. The unperturbed lesser and greater Green's functions can be obtained from Eqs. \eqref{eq4} and \eqref{eq5}.

The $2 \times 2$ Green's function above only gives one pole, whose energy is the Shiba energy
\begin{equation}
\epsilon_0 = \text{sgn}(\alpha) \Delta \frac{1-\alpha^2+\beta^2}{\sqrt{(1-\alpha^2+\beta^2)^2+4\alpha^2}}.
\label{eq_12}
\end{equation}

The quantum phase transition (QPT, $\epsilon_0=0$) happens at $\alpha=\sqrt{1+\beta^2}$. If $0<\alpha<\sqrt{1+\beta^2}$, which is below the QPT (weak coupling), then $\epsilon_0>0$. If $\alpha>\sqrt{1+\beta^2}$, which is above the QPT (strong coupling), then $\epsilon_0<0$. Changing the sign of $\alpha$, which means flipping the spin of the impurity, also changes the sign of $\epsilon_0$.

We are interested in the Green's function near $\omega = \epsilon_0$, in which the Green's function of the YSR can be simplified to
\begin{equation}
g_\mathrm{YSR}(\omega)=\frac{1}{\omega-\epsilon_0}\begin{pmatrix}
u^2 & uv \\
uv & v^2
\end{pmatrix},\label{eq_YSR_simple}
\end{equation}
with
\begin{equation}
u^2,v^2=\frac{2|\alpha|\pi \nu_0 \Delta(1+(\alpha\pm\beta)^2)}{((1-\alpha^2+\beta^2)^2+4\alpha^2)^{3/2}}.
\label{eq_13}
\end{equation}
Here, changing the sign of $\alpha$ exchanges $u^2$ and $v^2$.

Now we write the retarded and advanced Green's function
\begin{equation}
g_\mathrm{YSR}^{r,a}(\omega)=\frac{1}{\omega\pm i\Gamma-\epsilon_0}\begin{pmatrix}
u^2 & uv \\
uv & v^2
\end{pmatrix},
\end{equation}
in which the phenomenological parameter $\Gamma$ stands for the intrinsic relaxation of the Shiba state. With Eqs. \eqref{eq4} and \eqref{eq5} in mind, we can write the lesser and greater Green's functions as well as the densities of states:
\begin{align}
g_\mathrm{YSR}^{+-}(\omega)&=\frac{2i\Gamma f(\omega)}{(\omega-\epsilon_0)^2+\Gamma^2}\begin{pmatrix}
u^2 & uv \\
uv & v^2
\end{pmatrix},\label{eq_YSR_pm}\\
g_\mathrm{YSR}^{-+}(\omega)&=\frac{-2i\Gamma (1-f(\omega))}{(\omega-\epsilon_0)^2+\Gamma^2}\begin{pmatrix}
u^2 & uv \\
uv & v^2
\end{pmatrix},\label{eq_YSR_mp}\\
\rho_\mathrm{YSR,11}(\omega)&=\frac{1}{\pi}\frac{u^2\Gamma}{(\omega-\epsilon_0)^2+\Gamma^2},\:
\rho_\mathrm{YSR,22}(\omega)=\frac{1}{\pi}\frac{v^2\Gamma}{(\omega-\epsilon_0)^2+\Gamma^2}.\label{eq_YSR_dos}
\end{align}

\subsection{Thermal-direct ratio in YSR-BCS tunneling}
We can simluate the conventional YSR-BCS tunneling spectrum by plugging the BCS Green's function as one electrode ($g_R$) and the YSR Green's function as another electrode ($g_L$) into the expression for the tunneling current Eq. \eqref{eq6.2}. We know that $g_\mathrm{BCS}^{+-}(\omega)=2i \mathrm{Im}[g^a_\mathrm{BCS}(\omega)]f(\omega), g_\mathrm{BCS}^{-+}(\omega)=-2i \mathrm{Im}[g^a_\mathrm{BCS}(\omega)](1-f(\omega))$. It can also be shown that the density of states of a clean BCS superconductor $\rho_\mathrm{BCS}(\omega)=(1/\pi)\mathrm{Im}[g^a_\mathrm{BCS,11}(\omega)]=(1/\pi)\mathrm{Im}[g^a_\mathrm{BCS,22}(\omega)]$ is an even function, which sharply peaks at $\omega=\pm\Delta$ with $\rho_\mathrm{BCS}(\pm\Delta) \cong \frac{\nu_0 }{2}\sqrt{\frac{\Delta}{\eta}}$ for small $\eta$, which are the coherence peaks in the spectral function. Hence Eq. \eqref{eq6.2} becomes:
\begin{equation}
I_0=\frac{4\pi e|t|^2\Gamma}{h}\int\limits_{-\infty}^{+\infty} d\omega [u^2 \frac{ f(\omega)-f(\omega+\omega_0)}{(\omega-\epsilon_0)^2+\Gamma^2} \rho_\mathrm{BCS}(\omega+\omega_0)- v^2 \frac{ f(\omega)-f(\omega-\omega_0)}{(\omega-\epsilon_0)^2+\Gamma^2} \rho_\mathrm{BCS}(\omega-\omega_0)].\label{current_shiba_bcs_general}
\end{equation}

The function in the integral is only non-zero around $\omega\cong\epsilon_0$, so we substitute $x=\omega-\epsilon_0$ and limit the integration to $x\in [-L,L]$, where the parameter $L$ is chosen such that $\Gamma \ll L \ll k_\mathrm{B}T$ so that the Fermi function can be approximated as constant in the integration range. The existence of such $h$ is ensured by the assumption of $\Gamma \ll k_\mathrm{B}T \ll \Delta$ which is generally true for our experiment. Under this assumption, $f(\Delta)\cong 0, f(-\Delta)\cong 1$. We assume also $\epsilon_0>0$ for convenience. Given that $\rho_\mathrm{BCS}(\omega)$ sharply peaks at $\omega=\pm\Delta$, the first term in Eq.\ \eqref{current_shiba_bcs_general} corresponds to the peaks at $\omega_0=-\epsilon_0\pm\Delta$ ($t_+$ and $d_-$, where $t,d$ stand for thermal and direct YSR-BCS tunneling, and $+,-$ stand for the process at positive and negative voltage) and the second term corresponds to the peaks at $\omega_0=\epsilon_0\mp\Delta$ ($t_-$ and $d_+$). For $\epsilon_0$ not too close to $\Delta$ nor to zero (not within $\pm\eta$, which also holds generally in the experiment) so that at each resonance only one of the two terms is dominant, the expressions at the resonances become
\begin{align}
I_\mathrm{t+}&=I_0|_{\omega_0=\Delta-\epsilon_0}\cong\frac{4\pi e|t|^2\Gamma}{h}u^2\int_{-L}^{L} dx\frac{ f(\epsilon_0)-f(\Delta)}{x^2+\Gamma^2} \rho_\mathrm{BCS}(x+\Delta)
=u^2 f(\epsilon_0) C,\\
I_\mathrm{d-}&=I_0|_{\omega_0=-\Delta-\epsilon_0}\cong\frac{4\pi e|t|^2\Gamma}{h}u^2\int_{-L}^{L} dx\frac{ f(\epsilon_0)-f(-\Delta)}{x^2+\Gamma^2} \rho_\mathrm{BCS}(x-\Delta)
=u^2 (f(\epsilon_0)-1) C,\\
I_\mathrm{t-}&=I_0|_{\omega_0=-\Delta+\epsilon_0}\cong-\frac{4\pi e|t|^2\Gamma}{h}v^2\int_{-L}^{L} dx\frac{ f(\epsilon_0)-f(\Delta)}{x^2+\Gamma^2} \rho_\mathrm{BCS}(x+\Delta)
=-v^2 f(\epsilon_0) C,\\
I_\mathrm{d+}&=I_0|_{\omega_0=\Delta+\epsilon_0}\cong-\frac{4\pi e|t|^2\Gamma}{h}v^2\int_{-L}^{L} dx\frac{ f(\epsilon_0)-f(-\Delta)}{x^2+\Gamma^2} \rho_\mathrm{BCS}(x-\Delta)
=-v^2 (f(\epsilon_0)-1) C,\\
C&=\frac{4\pi e|t|^2\Gamma}{h}\int_{-L}^{L} dx\frac{1}{x^2+\Gamma^2} \rho_\mathrm{BCS}(x+\Delta),
\end{align}
where we utilize the even property of $\rho_\mathrm{BCS}(x)$ such that $\int_{-L}^{L} dx\frac{1}{x^2+\Gamma^2} \rho_\mathrm{BCS}(x-\Delta)=\int_{L}^{-L} d(-x)\frac{1}{(-x)^2+\Gamma^2} \rho_\mathrm{BCS}(-x-\Delta)=\int_{-L}^{L} dx\frac{1}{x^2+\Gamma^2} \rho_\mathrm{BCS}(x+\Delta)$.

It can be easily seen that the asymmetry of the thermal YSR-BCS peaks is reversed compared to the asymmetry of the direct peaks, $\frac{I_\mathrm{t+}}{I_\mathrm{t-}}=\frac{I_\mathrm{d-}}{I_\mathrm{d+}}=\frac{-u^2}{v^2}$, which was also shown in Ref. \cite{si_Ruby2015a}. The thermal-direct ratio $r_\mathrm{YSR-BCS}$ is
\begin{equation}
    r_\mathrm{YSR-BCS} = \left|\frac{I_\mathrm{t+}}{I_\mathrm{d-}}\right|=\left|\frac{I_\mathrm{t-}}{I_\mathrm{d+}}\right|=\left|\frac{f(\epsilon_0)}{f(\epsilon_0)-1}\right|=e^{-\epsilon_0/{k_\mathrm{B}T}},
\end{equation}
which is the Boltzmann factor. Here we use the current peak to calculate the ratio, but the $dI/dV$ peak height can also be shown to yield the same ratio. Consequently, we can measure the thermal-direct ratio of a conventional YSR-BCS tunenling process ($dI/dV$ or current peak), and the ratio should be the Boltzmann factor of the YSR state. Experimentally this is done by moving the YSR tip to a clean position on the surface without sample YSR states and measure the spectrum at low conductance, and use a clean tip to measure the intrinsic YSR state on the surface. In this way, the Boltzmann factor of the tip and sample YSR state can be separately characterized experimentally. This can also be used as a temperature sensor because $\epsilon_0$ can be easily read out in the spectrum, leaving $T$ as the only parameter left. We actually used this property to calibrate the temperature sensor at this temperature range of the machine. Notice that below approximately 0.6\,K, the thermal process is usually difficult to detect experimentally.

\subsection{Shiba-Shiba tunneling in the $2\times 2$ Nambu Space}

Here, we first demonstrate the Shiba-Shiba tunneling in the $2\times 2$ formalism established in the previous part, to understand the lineshape as well as some basic properties of Shiba-Shiba tunneling. This formalism is relatively simple, but it is limited in the way that the two YSR states can only have parallel or anti-parallel spin. There is only one pole in the YSR Green's function, which means that only one among the two processes (thermal or direct Shiba-Shiba) can exist in one spectrum. A $4\times 4$ formalism is needed to overcome this limitation and reflect more general experimental situations which will be discussed later.

To derive a current expression in the $2\times 2$ formalism, we need to insert two Shiba Green's functions as the left and right electrode. It is sufficient to use the simplified formula in Eq. \eqref{eq_YSR_simple}, because the Shiba-Shiba peak is always separated from other spectral features and thus only the Green's function near the resonance participates:
\begin{equation}
g_{L,R}^{r,a}(\omega)=\frac{1}{\omega\pm i\Gamma_{L,R}-\epsilon_{L,R}}\begin{pmatrix}
u_{L,R}^2 & u_{L,R}v_{L,R} \\
u_{L,R}v_{L,R} & v_{L,R}^2
\end{pmatrix}.
\end{equation}

Plugging Eq.\ \eqref{eq_YSR_dos} into Eq.\ \eqref{eq6.2}, we have

\begin{align}
I_0=\frac{4e|t|^2 \Gamma_L\Gamma_R}{h} \int_{-\infty}^{+\infty} d\omega&\left(u_L^2 u_R^2\frac{1}{(\omega-\omega_0-\epsilon_L)^2+\Gamma_L^2} \frac{1}{(\omega-\epsilon_R)^2+\Gamma_R^2}  \left(f(\omega-\omega_0)-f(\omega)\right)\right.\nonumber\\
&\left.+v_L^2 v_R^2\frac{1}{(\omega-\epsilon_L)^2+\Gamma_L^2} \frac{1}{(\omega-\omega_0-\epsilon_R)^2+\Gamma_R^2}  \left(f(\omega-\omega_0)-f(\omega)\right) \right),
\label{eq20}
\end{align}
which is a convolution of two Lorentzians and the Fermi function. This integral can be carried out analytically under the assumption
\begin{equation}
    \Gamma_{L,R} \ll k_\mathrm{B}T,
    \label{eq21}
\end{equation}
which is generally true for Shiba-Shiba tunneling even at 10\,mK \cite{si_Huang2020tunneling}.

The first term in the integral of Eq. \eqref{eq20} is only nonzero near $\omega = \epsilon_R, \omega_0 = \epsilon_R-\epsilon_L$ within a range of the order of $\Gamma_{L,R}$. The second term is only nonzero near $\omega = \epsilon_L, \omega_0 = \epsilon_L-\epsilon_R$ within a range on the order of $\Gamma_{L,R}$. Due to the assumption in Eq. \eqref{eq21}, the Fermi function can be approximated as constant in this region and taken out of the integral. Using the definite integral of two Lorentzian functions
\begin{equation}
\int\limits_{-\infty}^{+\infty}dx \frac{1}{(x-a_1)^2+g_1^2} \frac{1}{(x-a_2)^2+g_2^2}=\frac{\pi}{g_1g_2}\frac{g_1+g_2}{(a_1-a_2)^2+(g_1+g_2)^2},
\end{equation}

we arrive at the following expressions
\begin{align}
    I_0&=I_0^1+I_0^2,\\
    I_0^1&=\frac{4\pi e|t|^2}{h}u_L^2 u_R^2\left(f(\epsilon_L)-f(\epsilon_R)\right)\frac{\Gamma_L+\Gamma_R}{\left(\omega_0-(\epsilon_R-\epsilon_L)\right)^2+(\Gamma_L+\Gamma_R)^2},\label{eq22}\\
    I_0^2&=\frac{4\pi e|t|^2}{h}v_L^2 v_R^2\left(f(\epsilon_R)-f(\epsilon_L)\right)\frac{\Gamma_L+\Gamma_R}{\left(\omega_0-(\epsilon_L-\epsilon_R)\right)^2+(\Gamma_L+\Gamma_R)^2}.\label{eq23}
\end{align}

$I_0^1$ and $I_0^2$ have different signs of current and resonant voltage, meaning that they are responsible for the positive and negative Shiba-Shiba peaks. The current peak heights are
\begin{align}
    |I_0^1|_\mathrm{max}&=\frac{4\pi e|t|^2 |f(\epsilon_R)-f(\epsilon_L)|}{h(\Gamma_L+\Gamma_R)}u_L^2 u_R^2,\label{eq24}\\
    |I_0^2|_\mathrm{max}&=\frac{4\pi e|t|^2 |f(\epsilon_R)-f(\epsilon_L)|}{h(\Gamma_L+\Gamma_R)}v_L^2 v_R^2.\label{eq25}
\end{align}

The current peak areas are, however, independent of the intrinsic relaxation $\Gamma_{L,R}$:
\begin{align}
    A_1=\left|\int I_0^1 dV\right|&=\frac{4\pi^2 |t|^2 |f(\epsilon_R)-f(\epsilon_L)|}{h}u_L^2 u_R^2,\label{eq26}\\
    A_2=\left|\int I_0^2 dV\right|&=\frac{4\pi^2 |t|^2 |f(\epsilon_R)-f(\epsilon_L)|}{h}v_L^2 v_R^2.\label{eq27}
\end{align}
In the temperature-dependent measurement of direct Shiba-Shiba peak, the peak area is used rather than the peak height, because in this way the temperature dependence of the intrinsic relaxation can be ruled out, leaving only the possible variation of spin dynamics.

\subsection{Properties of Shiba-Shiba tunneling}

Continuing from the discussion before about the sign of $\alpha$ and the relation between $\alpha$ and $\sqrt{1+\beta^2}$, the situation of two YSR states with parallel spins (same side of QPT and parallel impurity spin or different side of QPT and opposite impurity spin) is equivalent to $\epsilon_L \epsilon_R>0$, and the situation of two YSR states with opposite spins (different side of QPT and parallel impurity spin or same side of QPT and opposite impurity spin) is equivalent to $\epsilon_L \epsilon_R<0$.

If $\epsilon_L \epsilon_R>0$, $I_0$ is exponentially suppressed at low temperature by the prefactor $f(\epsilon_L)-f(\epsilon_R)$, which refers to a \textit{thermal Shiba-Shiba process}. The bias voltage needed is $eV=\omega_0=\pm(|\epsilon_R|-|\epsilon_L|)$. The direct Shiba-Shiba process is missing in this situation, corresponding to $\theta=0^\circ$ in the main text.

If $\epsilon_L \epsilon_R<0$, $|f(\epsilon_L)-f(\epsilon_R)|$ approaches one at zero temperature, resulting in a \textit{direct Shiba-Shiba process}. The bias voltage needed is $eV=\omega_0=\pm(|\epsilon_R|+|\epsilon_L|)$. The thermal Shiba-Shiba process is missing in this situation, corresponding to $\theta=180^\circ$ in the main text.

Note that due to the limitation of the $2\times 2$ formalism, the spins are either parallel or anti-parallel, so that the thermal Shiba-Shiba and direct Shiba-Shiba processes cannot coexist. To account for arbitrary angles, we need to extend to the $4\times 4$ formalism as discussed in the following.

\subsection{Extension to the 4$\times$4 Nambu space}

To account for arbitrary angles between tip and sample YSR spins, we extend the Green's function theory to the $4 \times 4$ spin-Nambu space. With the convention used in Ref. \cite{si_Shiba1968}, the basis for the $4 \times 4$ spin-Nambu space is
\begin{align}
A_k &= \begin{pmatrix}
       a_{k\uparrow} \\
       a_{k\downarrow} \\
       a_{-k\uparrow}^\dagger \\
       a_{-k\downarrow}^\dagger
     \end{pmatrix}
,\quad A_k^\dagger=(a_{k\uparrow}^\dagger,a_{k\downarrow}^\dagger,a_{-k\uparrow},a_{-k\downarrow}).\label{eq_operator}
\end{align}

The general expression of the current can be extended from Eq. \eqref{eq_current_ori} to the $4 \times 4$ spin-Nambu space \cite{si_Villas2020}:
\begin{equation}
    I(\tau)=\frac{e}{2\hbar}\mathrm{Tr}[(\tau_3 \otimes\sigma_0)(\hat{t}(\tau)G_{RL}^{+-}(\tau,\tau)-G_{LR}^{+-}(\tau,\tau)\hat{t}^{\dagger}(\tau))],
    \label{eq_current_4by4_general}
\end{equation}
where
\begin{equation}
 \tau_3 \otimes\sigma_0=\begin{pmatrix}
1 & 0 & 0 & 0  \\
0 & 1 & 0 & 0  \\
0 & 0 & -1 & 0  \\
0 & 0 & 0 & -1
\end{pmatrix}\quad\text{and}\quad\hat{t}(\tau)=\begin{pmatrix}
te^{i\omega_0 \tau/\hbar} & 0 & 0 & 0  \\
0 & te^{i\omega_0 \tau/\hbar} & 0 & 0  \\
0 & 0 & -t^*e^{-i\omega_0 \tau/\hbar} & 0  \\
0 & 0 & 0 & -t^*e^{-i\omega_0 \tau/\hbar}
\end{pmatrix}.
\end{equation}
Here, $\tau_i$ and $\sigma_i$ (i= 1,2,3) stand for Pauli matrices in Nambu and spin space, respectively, with $\tau_0$ and $\sigma_0$ being the unit matrices in the corresponding spaces.

From similar derivations as in the $2\times 2$ case (Eq. \eqref{eq6}), we can write the current in the linear regime:
\begin{align}
I_0=\frac{e|t|^2}{2h}\int_{-\infty}^{+\infty} &d\omega [g_{LL,11}^{+-}(\omega)g_{RR,11}^{-+}(\omega+\omega_0)+g_{LL,22}^{+-}(\omega)g_{RR,22}^{-+}(\omega+\omega_0)-g_{LL,33}^{+-}(\omega)g_{RR,33}^{-+}(\omega-\omega_0)-g_{LL,44}^{+-}(\omega)g_{RR,44}^{-+}(\omega-\omega_0)\nonumber\\
&-g_{RR,11}^{+-}(\omega)g_{LL,11}^{-+}(\omega-\omega_0)-g_{RR,22}^{+-}(\omega)g_{LL,22}^{-+}(\omega-\omega_0)+g_{RR,33}^{+-}(\omega)g_{LL,33}^{-+}(\omega+\omega_0)+g_{RR,44}^{+-}(\omega)g_{LL,44}^{-+}(\omega+\omega_0)].\label{eq_current_4by4}
\end{align}

Similarly, we can write in terms of the density of states $\rho_{ii}(\omega)=\mp\frac{1}{\pi}\Im{[g^{r,a}_{ii}(\omega)]},i\in \{1,2,3,4\}$:
\begin{equation}
\begin{split}
I_0&=\frac{2\pi^2 e|t|^2}{h}\int_{-\infty}^{+\infty}  \big(\rho_{L,11}(\omega)\rho_{R,11}(\omega+\omega_0)+\rho_{L,22}(\omega)\rho_{R,22}(\omega+\omega_0)\\
&+\rho_{L,33}(\omega+\omega_0)\rho_{R,33}(\omega)+\rho_{L,44}(\omega+\omega_0)\rho_{R,44}(\omega)\big)\big(f(\omega)-f(\omega+\omega_0)\big) d\omega.
\label{eq_current_4by4.2}
\end{split}
\end{equation}

\subsubsection{Arbitrary spin angle in the Green's function}

To account for arbitrary angles of the impurity spin $S$ away from the quantization axis set in Eq. \eqref{eq_operator}, we will need the general rotation matrix $\hat{R}$ to rotate the basis $A_k^\dagger= (a_{k\uparrow}^\dagger,\ a_{k\downarrow}^\dagger,\ a_{-k\uparrow},\ a_{-k\downarrow})$:

\begin{equation}
\hat{R}=\begin{pmatrix}
    \cos{\frac{\theta}{2}} & -\sin{\frac{\theta}{2}} & 0 & 0 \\
    \sin{\frac{\theta}{2}} & \cos{\frac{\theta}{2}} & 0 & 0 \\
    0 & 0 & \cos{\frac{\theta}{2}} & -\sin{\frac{\theta}{2}} \\
    0 & 0 & \sin{\frac{\theta}{2}} & \cos{\frac{\theta}{2}}
\end{pmatrix},
\end{equation}
where $\hat{R}\hat{R}^\mathrm{T}=\mathbb{1}$ with $\theta \in [0,\pi]$ being the angle relative to spin $\uparrow$. There are two equivalent methods. One obvious option is to directly rotate the basis of the Hamiltonian such that $\widetilde{H}=\hat{R}H\hat{R}^\mathrm{T}$. The new Green's functions will be
\begin{equation}
\widetilde{g}=(\omega-\widetilde{H})^{-1}=(\omega-\hat{R}H\hat{R}^\mathrm{T})^{-1}=(\hat{R}\omega \hat{R}^\mathrm{T}-\hat{R}H\hat{R}^\mathrm{T})^{-1}=(\hat{R}^\mathrm{T})^{-1}(\omega -H)^{-1}\hat{R}^{-1}=\hat{R}g\hat{R}^\mathrm{T}.
\label{eq_4by4_rotate_g}
\end{equation}

Now consider the tunneling current after the spin rotation. We start with two general Green's functions in the main spin quantization axis, which can be block-diagonalized in the absence of spin-orbit coupling and other spin-flip scattering terms in the Hamiltonian \cite{si_Zhu2016}, which is generally true also for YSR states (see Eq.\ \eqref{eq_ysr_4by4_main}), resulting in the zero terms in the matrices
\begin{equation}
g_{L}=\begin{pmatrix}
    g_{L,11} & 0 & 0 & g_{L,14} \\
    0 & g_{L,22} & g_{L,23} & 0 \\
    0 & g_{L,32} & g_{L,33} & 0 \\
    g_{L,41} & 0 & 0 & g_{L,44}
\end{pmatrix},
g_{R}=\begin{pmatrix}
    g_{R,11} & 0 & 0 & g_{R,14} \\
    0 & g_{R,22} & g_{R,23} & 0 \\
    0 & g_{R,32} & g_{R,33} & 0 \\
    g_{R,41} & 0 & 0 & g_{R,44}
\end{pmatrix}.
\end{equation}

Now we rotate $g_R$ to $\widetilde{g_R}=\hat{R}g_R\hat{R}^\mathrm{T}$ and plug $g_L$ and $\widetilde{g_R}$ into Eq. \eqref{eq_current_4by4.2}, we get
\begin{equation}
\begin{split}
I_0=\frac{2\pi^2 e|t|^2}{h}\sum_{\sigma\in\{\uparrow,\downarrow\}}\int_{-\infty}^{+\infty}  \big(\rho^e_{L,\sigma}(\omega)\rho^e_{R,\sigma}(\omega+\omega_0)\cos^2{\frac{\theta}{2}}+\rho^e_{L,\sigma}(\omega)\rho^e_{R,\Bar{\sigma}}(\omega+\omega_0)\sin^2{\frac{\theta}{2}}\\
+\rho^h_{L,\sigma}(\omega+\omega_0)\rho^h_{R,\sigma}(\omega)\cos^2{\frac{\theta}{2}}+\rho^h_{L,\sigma}(\omega+\omega_0)\rho^h_{R,\Bar{\sigma}}(\omega)\sin^2{\frac{\theta}{2}}\big)\big(f(\omega)-f(\omega+\omega_0)\big) d\omega,
\end{split}
\end{equation}
where $\rho^e_\uparrow=\rho_{11},\rho^e_\downarrow=\rho_{22},\rho^h_\uparrow=\rho_{33},\rho^h_\downarrow=\rho_{44}$ are the densities of states coming from the corresponding terms in the non-rotated Green's function matrices $g_{L,R}$ and $\Bar{\sigma}$ denotes the opposite spin of $\sigma$. If we combine the electron part and the hole part, the above formula becomes Eq.\ (1) in the main text, which is also the ordinary spin-polarized tunneling formula.

\subsubsection{YSR Green's function in the main quantization axis}
We first deal with the situation where there is no rotation with respect to the main quantization basis. The Hamiltonian is still Eq.\ \eqref{eq_Hamil}, but as a $4\times 4$ matrix
\begin{equation}
\begin{split}
    H&=H_\mathrm{SC}+H_\mathrm{J}+H_\mathrm{U},\\
    H_\mathrm{SC}&=\begin{pmatrix}
    \xi_k & 0 & 0 & \Delta \\
    0 & \xi_k & -\Delta & 0 \\
    0 & -\Delta & -\xi_k & 0 \\
    \Delta & 0 & 0 & -\xi_k
    \end{pmatrix},\quad
    H_\mathrm{J}=JS\begin{pmatrix}
    1 & 0 & 0 & 0 \\
    0 & -1 & 0 & 0 \\
    0 & 0 & -1 & 0 \\
    0 & 0 & 0 & 1
    \end{pmatrix},\quad
    H_\mathrm{U}=U\begin{pmatrix}
    1 & 0 & 0 & 0 \\
    0 & 1 & 0 & 0 \\
    0 & 0 & -1 & 0 \\
    0 & 0 & 0 & -1
    \end{pmatrix}.\label{eq_Hamil_4by4}
\end{split}
\end{equation}

Plugging the Hamiltonian in the formula $g=(\omega-H)^{-1}$ and using the same substitution $\alpha=\pi\nu_0 JS$ and $\beta=\pi\nu_0 V$, we obtain the YSR Green's function in the $4 \times 4$ spin-Nambu space
\begin{align}
g_\mathrm{YSR}&=\begin{pmatrix}
C_2\Big(\omega-(\alpha-\beta)\sqrt{\Delta^2-\omega^2}\Big) & 0 & 0 & C_2\Delta \\
0 & C_1\Big(\omega+(\alpha+\beta)\sqrt{\Delta^2-\omega^2}\Big) & -C_1\Delta & 0 \\
0 & -C_1\Delta & C_1\Big(\omega+(\alpha-\beta)\sqrt{\Delta^2-\omega^2}\Big) & 0 \\
C_2\Delta & 0 & 0 & C_2\Big(\omega-(\alpha+\beta)\sqrt{\Delta^2-\omega^2}\Big)
\end{pmatrix}\label{eq_ysr_4by4_main},\\
C_1&=\frac{-\pi\nu_0}{-2\alpha\omega+(1-\alpha^2+\beta^2)\sqrt{\Delta^2-\omega^2}},\\
C_2&=\frac{-\pi\nu_0}{2\alpha\omega+(1-\alpha^2+\beta^2)\sqrt{\Delta^2-\omega^2}}.
\end{align}

This reduces directly to the $2 \times 2$ case (Eq.\ \eqref{eq_YSR}) because the matrix can be block-diagonalized into two equivalent $2 \times 2$ Green's function $C_1\Big( \begin{smallmatrix}(\omega+(\alpha+\beta)\sqrt{\Delta^2-\omega^2}) & -\Delta\\ -\Delta & (\omega+(\alpha-\beta)\sqrt{\Delta^2-\omega^2})\end{smallmatrix}\Big)$ and $C_2\Big( \begin{smallmatrix}(\omega-(\alpha-\beta)\sqrt{\Delta^2-\omega^2}) & \Delta\\ \Delta & (\omega-(\alpha+\beta)\sqrt{\Delta^2-\omega^2})\end{smallmatrix}\Big)$. Now we have two poles rather than only one previously
\begin{equation}
\epsilon= \pm\epsilon_0,\; \epsilon_0=\Delta \frac{1-\alpha^2+\beta^2}{\sqrt{(1-\alpha^2+\beta^2)^2+4\alpha^2}}.
\label{eq_energy_4by4}
\end{equation}

\subsubsection{Rotating the YSR Green's function}

Plugging Eq. \eqref{eq_ysr_4by4_main} into Eq. \eqref{eq_4by4_rotate_g}, we can get the rotated YSR Green's function. From now on we let $\alpha \geq 0$ because the spin direction is contained in $\theta$. Since only diagonal terms are relevant for the current at low conductance (Eqs. \eqref{eq_current_4by4} and \eqref{eq_current_4by4.2}), we only write down those parts here
\begin{equation}
\begin{split}
G_\mathrm{YSR}&=\begin{pmatrix}
G_{11} &  &  &\ddots  \\
 & G_{22} &  &  \\
 &  & G_{33} &  \\
 &\ddots  &  & G_{44}
\end{pmatrix}, \\
    G_{11}&=C\Big((\beta-\alpha\cos{\theta})(\Delta^2-\omega^2)(1-\alpha^2+\beta^2)+\omega(1+\alpha^2+\beta^2-2\alpha\beta\cos{\theta})\sqrt{\Delta^2-\omega^2}-2\omega^2\alpha\cos{\theta}\Big),\\
    G_{22}&=C\Big((\beta+\alpha\cos{\theta})(\Delta^2-\omega^2)(1-\alpha^2+\beta^2)+\omega(1+\alpha^2+\beta^2+2\alpha\beta\cos{\theta})\sqrt{\Delta^2-\omega^2}+2\omega^2\alpha\cos{\theta}\Big),\\
    G_{33}&=C\Big(-(\beta-\alpha\cos{\theta})(\Delta^2-\omega^2)(1-\alpha^2+\beta^2)+\omega(1+\alpha^2+\beta^2-2\alpha\beta\cos{\theta})\sqrt{\Delta^2-\omega^2}+2\omega^2\alpha\cos{\theta}\Big),\\
    G_{44}&=C\Big(-(\beta+\alpha\cos{\theta})(\Delta^2-\omega^2)(1-\alpha^2+\beta^2)+\omega(1+\alpha^2+\beta^2+2\alpha\beta\cos{\theta})\sqrt{\Delta^2-\omega^2}-2\omega^2\alpha\cos{\theta}\Big),\\
    C&=\frac{-\pi\nu_0}{(\Delta^2-\omega^2)(1-\alpha^2+\beta^2)^2-4\alpha^2\omega^2}.
\end{split}\label{eq_ysr_rotated_general}
\end{equation}

\begin{figure}
    \centering
    \includegraphics[width=0.85\columnwidth]{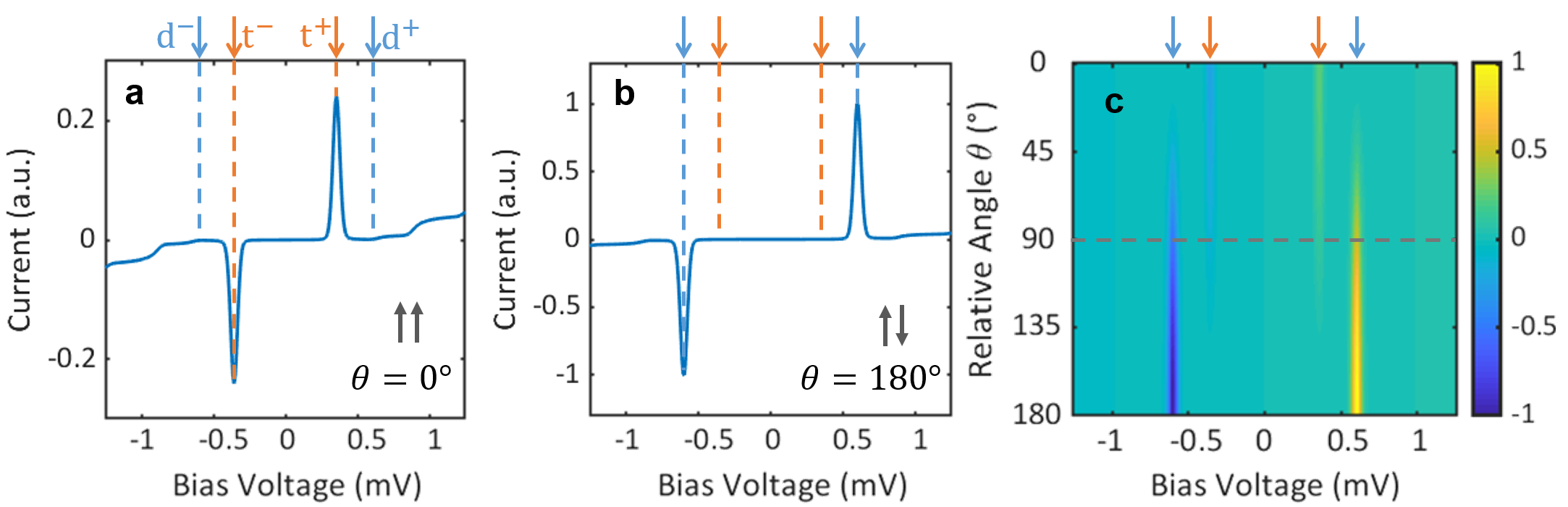}
    \caption{\textbf{Simulated I(V) spectra of Shiba-Shiba tunneling at different relative spin angles.} The spectra are calculated assuming a temperature of 1\,K in the case of (a) parallel spins, (b) anti-parallel spins and (c) arbitrary relative spin angles. The orange arrows and dashed lines ($t^{+,-}$) label the thermal Shiba-Shiba peaks, while the blue ones ($d^{+,-}$) label the direct Shiba-Shiba peaks. When the spins are parallel, only thermal Shiba-Shiba process is allowed. When the spins are anti-parallel, only direct Shiba-Shiba process is allowed. Intermediate relative angles result in the coexistence of both Shiba-Shiba processes.}
    \label{fig_si_3}
\end{figure}

The YSR energy as the poles of $G_\mathrm{YSR}$ (the denomenator of $C=0$) remains the same as in Eq. \eqref{eq_energy_4by4}, which is independent of $\theta$ as expected. Plugging Eq. \ref{eq_ysr_rotated_general} into Eq. \ref{eq_current_4by4.2}, we can directly simulate the I(V) (or dI/dV) spectra of Shiba-Shiba tunneling in the low conductance regime with arbitrary relative spin angles, shown in Fig. \ref{fig_si_3}, confirming the relevant discussion in the main text. Apart from this numerical calculation, we further focus on the Green's function near the YSR energy and derive the analytical formula of Shiba-Shiba peaks in the following part.

Near $\omega \rightarrow -\epsilon_0=-\Delta \frac{1-\alpha^2+\beta^2}{\sqrt{(1-\alpha^2+\beta^2)^2+4\alpha^2}}$, the Green's function can be simplified as
\begin{equation}
\begin{split}
    G_\mathrm{YSR}|_{\omega \rightarrow -\epsilon_0}&=\frac{1}{\omega+\epsilon_0}\begin{pmatrix}
u_{-\uparrow}^2 &  &  &\ddots  \\
 & u_{-\downarrow}^2 &  &  \\
 &  & v_{-\uparrow}^2 &  \\
 &\ddots  &  & v_{-\downarrow}^2
\end{pmatrix},\\
u_{-\uparrow}^2&=\frac{2\pi\nu_0\alpha\Delta}{((1-\alpha^2+\beta^2)^2+4\alpha^2)^{3/2}}(1+(\alpha-\beta)^2)\cos^2{\frac{\theta}{2}},\\
u_{-\downarrow}^2&=\frac{2\pi\nu_0\alpha\Delta}{((1-\alpha^2+\beta^2)^2+4\alpha^2)^{3/2}}(1+(\alpha-\beta)^2)\sin^2{\frac{\theta}{2}},\\
v_{-\uparrow}^2&=\frac{2\pi\nu_0\alpha\Delta}{((1-\alpha^2+\beta^2)^2+4\alpha^2)^{3/2}}(1+(\alpha+\beta)^2)\sin^2{\frac{\theta}{2}},\\
v_{-\downarrow}^2&=\frac{2\pi\nu_0\alpha\Delta}{((1-\alpha^2+\beta^2)^2+4\alpha^2)^{3/2}}(1+(\alpha+\beta)^2)\cos^2{\frac{\theta}{2}}.
\end{split}\label{eq_ysr_4by4_n}
\end{equation}

For $\theta=0$, which is the spin-parallel case, $u_{-\downarrow}^2=v_{-\uparrow}^2=0$ and $u_{-\uparrow}^2,v_{-\downarrow}^2$ exchanges $e^--h^+$ asymmetry in Eq. \eqref{eq_13}. For $\theta=\pi$, which is the spin-opposite case, $u_{-\uparrow}^2=v_{-\downarrow}^2=0$ and $u_{-\downarrow}^2, v_{-\uparrow}^2$ also exchanges the $e^--h^+$ asymmetry in Eq. \eqref{eq_13}.

Near $\omega \rightarrow +\epsilon_0=\Delta \frac{1-\alpha^2+\beta^2}{\sqrt{(1-\alpha^2+\beta^2)^2+4\alpha^2}}$, the Green's function can be simplified as:
\begin{equation}
\begin{split}
    G_\mathrm{YSR}|_{\omega \rightarrow +\epsilon_0}&=\frac{1}{\omega-\epsilon_0}\begin{pmatrix}
u_{+\uparrow}^2 &  &  &\ddots  \\
 & u_{+\downarrow}^2 &  &  \\
 &  & v_{+\uparrow}^2 &  \\
 &\ddots  &  & v_{+\downarrow}^2
\end{pmatrix},\\
u_{+\uparrow}^2&=\frac{2\pi\nu_0\alpha\Delta}{((1-\alpha^2+\beta^2)^2+4\alpha^2)^{3/2}}(1+(\alpha+\beta)^2)\sin^2{\frac{\theta}{2}},\\
u_{+\downarrow}^2&=\frac{2\pi\nu_0\alpha\Delta}{((1-\alpha^2+\beta^2)^2+4\alpha^2)^{3/2}}(1+(\alpha+\beta)^2)\cos^2{\frac{\theta}{2}},\\
v_{+\uparrow}^2&=\frac{2\pi\nu_0\alpha\Delta}{((1-\alpha^2+\beta^2)^2+4\alpha^2)^{3/2}}(1+(\alpha-\beta)^2)\cos^2{\frac{\theta}{2}},\\
v_{+\downarrow}^2&=\frac{2\pi\nu_0\alpha\Delta}{((1-\alpha^2+\beta^2)^2+4\alpha^2)^{3/2}}(1+(\alpha-\beta)^2)\sin^2{\frac{\theta}{2}}.
\end{split}\label{eq_ysr_4by4_p}
\end{equation}

The situation here is the opposite to that in $\omega\rightarrow -\epsilon_0$. For $\theta=0$, which is the spin-parallel case, $u_{+\uparrow}^2=v_{+\downarrow}^2=0$ and $u_{+\downarrow}^2, v_{+\uparrow}^2$ recovers the $e^--h^+$ asymmetry in the $2\times 2$ case (Eq. \eqref{eq_13}). For $\theta=\pi$, which is the spin-opposite case, $u_{+\downarrow}^2=v_{+\uparrow}^2=0$ and $u_{+\uparrow}^2,v_{+\downarrow}^2$ recovers Eq. \eqref{eq_13}.

\subsubsection{Shiba-Shiba tunneling with arbitrary spin angle}

To account for the Shiba-Shiba tunneling with arbitrary spin angles, there are in general two equivalent options. The first method is to rotate the Green's function of one electrode while keeping the other electrode untouched. Alternatively, we can leave the Hamiltonian and Green's functions untouched while absorbing the rotation into the tunneling matrix $\widetilde{\hat{t}}=\hat{t}\hat{R}$, which would be the second method. In the following, we will follow the first method.

We write the tip (parameters $\alpha_L>0, \beta_L, \Gamma_L$, Green's function in the main quantization axis) and sample (parameters $\alpha_R>0, \beta_R, \Gamma_R, \theta$, rotated Green's function) YSR state as:
\begin{equation}
\begin{split}
G_\mathrm{L}^{r,a}&=\frac{1}{\omega\pm i\Gamma_L-\epsilon_L}\begin{pmatrix}
u_{L+\uparrow}^2 &  &  &\ddots  \\
 & u_{L+\downarrow}^2 &  &  \\
 &  & v_{L+\uparrow}^2 &  \\
 &\ddots  &  & v_{L+\downarrow}^2
\end{pmatrix}+\frac{1}{\omega\pm i\Gamma_L+\epsilon_L}\begin{pmatrix}
u_{L-\uparrow}^2 &  &  &\ddots  \\
 & u_{L-\downarrow}^2 &  &  \\
 &  & v_{L-\uparrow}^2 &  \\
 &\ddots  &  & v_{L-\downarrow}^2
\end{pmatrix},\\
G_\mathrm{R}^{r,a}&=\frac{1}{\omega\pm i\Gamma_R-\epsilon_R}\begin{pmatrix}
u_{R+\uparrow}^2 &  &  &\ddots  \\
 & u_{R+\downarrow}^2 &  &  \\
 &  & v_{R+\uparrow}^2 &  \\
 &\ddots  &  & v_{R+\downarrow}^2
\end{pmatrix}+\frac{1}{\omega\pm i\Gamma_R+\epsilon_R}\begin{pmatrix}
u_{R-\uparrow}^2 &  &  &\ddots  \\
 & u_{R-\downarrow}^2 &  &  \\
 &  & v_{R-\uparrow}^2 &  \\
 &\ddots  &  & v_{R-\downarrow}^2
\end{pmatrix}.\label{eq_GR_4by4}
\end{split}
\end{equation}

After some algebra, we have

\begin{equation}
\begin{split}
I_0&=\frac{16\pi e|t|^2}{h}\Gamma \frac{\pi\nu_{0,L}\alpha_L\Delta_L}{((1-\alpha_L^2+\beta_L^2)^2+4\alpha_L^2)^{3/2}} \frac{\pi\nu_{0,R}\alpha_R\Delta_R}{((1-\alpha_R^2+\beta_R^2)^2+4\alpha_R^2)^{3/2}}\times\\
&\Big(\frac{1}{(\omega_0+\epsilon_L-\epsilon_R)^2+\Gamma^2}\big(1+(\alpha_L+\beta_L)^2\big)\big(1+(\alpha_R+\beta_R)^2\big)\cos^2{\frac{\theta}{2}}\big(f(\epsilon_L)-f(\epsilon_R)\big)\\
&+\frac{1}{(\omega_0+\epsilon_L+\epsilon_R)^2+\Gamma^2}\big(1+(\alpha_L+\beta_L)^2\big)\big(1+(\alpha_R-\beta_R)^2\big)\sin^2{\frac{\theta}{2}}\big(f(\epsilon_L)-f(-\epsilon_R)\big)\\
&+\frac{1}{(\omega_0-\epsilon_L-\epsilon_R)^2+\Gamma^2}\big(1+(\alpha_L-\beta_L)^2\big)\big(1+(\alpha_R+\beta_R)^2\big)\sin^2{\frac{\theta}{2}}\big(f(-\epsilon_R)-f(\epsilon_L)\big)\\
&+\frac{1}{(\omega_0-\epsilon_L+\epsilon_R)^2+\Gamma^2}\big(1+(\alpha_L-\beta_L)^2\big)\big(1+(\alpha_R-\beta_R)^2\big)\cos^2{\frac{\theta}{2}}\big(f(\epsilon_R)-f(\epsilon_L)\big)
\Big).
\end{split}\label{eq_4by4_final}
\end{equation}

The four terms in Eq. \eqref{eq_4by4_final} stand for positive and negative peaks of thermal and direct Shiba-Shiba processes. For the convenience of the following discussion we assume $\epsilon_L>0,  \epsilon_R>0$. Therefore, $\omega_0=\pm(\epsilon_L-\epsilon_R)$ are the thermal peaks and $\omega_0=\pm(\epsilon_L+\epsilon_R)$ are the direct peaks. Now we plug in the four terms of Eq. \eqref{eq_4by4_final} into the thermal-direct ratio $r$ and assume low temperature

\begin{equation}
\begin{split}
    r=\sqrt{\frac{p_\mathrm{t+}p_\mathrm{t-}}{p_\mathrm{d+}p_\mathrm{d-}}}&=\left|
    \frac{\cos^2{\frac{\theta}{2}}}{\sin^2{\frac{\theta}{2}}}\times\frac{f(\epsilon_L)-f(\epsilon_R)}{f(\epsilon_L)-f(-\epsilon_R)}\right|\\
    &=\left|
    \cot^2{\frac{\theta}{2}}\times\frac{e^{-\epsilon_L/k_\mathrm{B}T}-e^{-\epsilon_R/k_\mathrm{B}T}}{1-e^{-\epsilon_L/k_\mathrm{B}T}e^{-\epsilon_R/k_\mathrm{B}T}}\right|\\
    &\cong \cot^2{\frac{\theta}{2}}\times r_\mathrm{th},\\
    r_\mathrm{th}&=\left|e^{-\epsilon_L/k_\mathrm{B}T}-e^{-\epsilon_R/k_\mathrm{B}T}\right|,
    \label{eq_ratio_4by4}
\end{split}
\end{equation}
where $p_\mathrm{t+,t-}$ stand for the pair of thermal Shiba-Shiba current peaks (either peak height or peak area) and $p_\mathrm{d+,d-}$ stand for direct Shiba-Shiba peaks, with $+,-$ denoting the peak at positive and negative voltage, consistent with the notation in the main text. This definition of $r$ eliminates the coherence factors $u^2$ and $v^2$ (i.e.,\ intensity and asymmetry) from the equation and makes $r$ only depend on $\theta$ and the Boltzmann factors, reducing the ambiguity of experimental interpretation.

For $\theta=90^\circ, r_{90^\circ}=r_\mathrm{th}$. If one of the spins is freely rotating, the thermal-direct Shiba-Shiba ratio $r_\mathrm{free}$ is
\begin{equation}
\begin{split}
    r_\mathrm{free}&=\left|\frac{\frac{1}{4\pi}\int_0^{\pi}2\pi \sin{\theta}(1+\cos{\theta})d\theta}{\frac{1}{4\pi}\int_0^{\pi}2\pi \sin{\theta}(1-\cos{\theta})d\theta}\right|\times r_\mathrm{th}=r_\mathrm{th}.
\end{split}
\end{equation}

\section{Requirements for a topological chain of YSR states}

The model of a one dimensional topological superconductor due to Kitaev is described by the following Hamiltonian
\begin{align}
H_\text{chain}= \sum_{n=1}^N E_{Sn} \gamma_n^\dagger\gamma_n+\sum_{n=1}^{N-1}\left(\tau_d\gamma_n^\dagger\gamma_{n+1}^\dagger+\tau_t \gamma_n^\dagger\gamma_{n+1}+\text{h.c.}\right).
\end{align}
This model consists of a chain of $N$ fermionic sites, each characterized by energy $E_{Sn}$ and corresponding creation (annihilation) operators given by $\gamma_n^\dagger$ ($\gamma_n$), for $n=\{1,\ldots,N\}$. The chain has two phases, a trivial phase where the
spectrum has a gap and no edge modes are present, and a topological phase where the spectrum contains a single fermionic state at zero energy formed by combining two Majorana modes localized at the two edges of the chain.

The topological phase of the chain is sensitive to the two amplitudes $\tau_d$ and $\tau_t$ describing coupling between nearest neighbor sites. The amplitude $\tau_d$ describes the pair creation process, while $\tau_t$ is the amplitude of the tunnel hopping process. When both amplitudes exceed the energies of the sites, $\tau_d,\tau_t\gg E_{Sn}$ for all sites, the chain is in the topological phase. In contrast, when either $\tau_d < E_{Sn}$, or $\tau_t < E_{Sn}$, the topological phase is lost and the chain does not host Majorana modes at its edges.

The above Hamiltonian can be realized by a chain of tunnel coupled spin-polarized YSR states on a common superconducting substrate \cite{si_Nadj2013,si_Pientka2013,si_Pientka2014,si_Li2014,si_Heimes2014,si_Hoffman2016,si_Poyhonen2016,si_Andolina2017,si_Theiler2019}, each with energy $E_{Sn}$. If the dynamics of the magnetic impurity spins
are decoupled from the electronic degrees of freedom, the above Hamiltonian describes the low energy excitations. In the analogy between the tunnel junction and the YSR chain, $\tau_d$ is the amplitude of the direct Shiba-Shiba tunneling process and $\tau_t$ the amplitude of the thermal Shiba-Shiba tunneling process. The YSR chain will reach the topological phase as long as both amplitudes exceed a certain threshold set by the disorder of the individual Shiba energies. It is, therefore, important to show that the relative spin-polarization of neighboring YSR states allows the two processes to coexist. The mechanism leading to this coexistence in the case of the experiment Ref.~\onlinecite{si_Nadj-Perge2014} remains under theoretical debate. A number of works have proposed that the RKKY interaction in such a chain leads to a helical ordering \cite{si_Klinovaja2013,si_Vazifeh2013,si_Braunecker2013,si_Pientka2013,si_Pientka2014,si_Hoffman2016,si_Andolina2017}, while a counter view holds that anisotropy due to the superconducting surface may lead to ferromagnetic ordering, where the spin blockade of the pair creation process is lifted only if the chain exhibits strong spin-orbit coupling \cite{si_Li2014,si_Poyhonen2016,si_Theiler2019}. To shed light on these discussions, separately detecting the two processes between neighboring sites experimentally and clarifying the role of spin there is desired.

Our results demonstrate the coexistence of the two processes for a pair of YSR impurities, one placed on the substrate and one on the tip of the STM, which paves the road for further research regarding the conditions the relative spin orientation in a chain of atomic impurities, such as realized in Ref.~\onlinecite{si_Nadj-Perge2014, si_Jack2019}, exhibits the same lack of anisotropy. Our results allow us to conclude that a functionalized STM tip with an YSR state will couple to an impurity chain in such a way that the topological phase of the chain can be extended to the YSR state on the tip.


\begin{thebibliography}{10}
\expandafter\ifx\csname url\endcsname\relax
  \def\url#1{\texttt{#1}}\fi
\expandafter\ifx\csname urlprefix\endcsname\relax\def\urlprefix{URL }\fi
\providecommand{\bibinfo}[2]{#2}
\providecommand{\eprint}[2][]{\url{#2}}

\bibitem{Pan2000}
\bibinfo{author}{Pan, S.~H.} \emph{et~al.}
\newblock \bibinfo{title}{Imaging the effects of individual zinc impurity atoms
  on superconductivity in {Bi$_2$Sr$_2$CaCu$_2$O$_{8+\delta}$}}.
\newblock \emph{\bibinfo{journal}{Nature}} \textbf{\bibinfo{volume}{403}},
  \bibinfo{pages}{746--750} (\bibinfo{year}{2000}).

\bibitem{Nadj-Perge2014}
\bibinfo{author}{Nadj-Perge, S.} \emph{et~al.}
\newblock \bibinfo{title}{Observation of {Majorana} fermions in ferromagnetic
  atomic chains on a superconductor}.
\newblock \emph{\bibinfo{journal}{Science}} \textbf{\bibinfo{volume}{346}},
  \bibinfo{pages}{602--607} (\bibinfo{year}{2014}).

\bibitem{Menard2017}
\bibinfo{author}{M{\'e}nard, G.~C.} \emph{et~al.}
\newblock \bibinfo{title}{Two-dimensional topological superconductivity in
  {Pb/Co/Si(111)}}.
\newblock \emph{\bibinfo{journal}{Nat. Commun.}} \textbf{\bibinfo{volume}{8}},
  \bibinfo{pages}{2040} (\bibinfo{year}{2017}).

\bibitem{Kezilebieke2019}
\bibinfo{author}{Kezilebieke, S.}, \bibinfo{author}{Zitko, R.},
  \bibinfo{author}{Dvorak, M.}, \bibinfo{author}{Ojanen, T.} \&
  \bibinfo{author}{Liljeroth, P.}
\newblock \bibinfo{title}{Observation of coexistence of {Yu-Shiba-Rusinov}
  states and spin-flip excitations}.
\newblock \emph{\bibinfo{journal}{Nano Lett.}} \textbf{\bibinfo{volume}{19}},
  \bibinfo{pages}{4614--4619} (\bibinfo{year}{2019}).

\bibitem{Odobesko2020}
\bibinfo{author}{Odobesko, A.} \emph{et~al.}
\newblock \bibinfo{title}{Observation of tunable single-atom {Yu-Shiba-Rusinov}
  states}.
\newblock \emph{\bibinfo{journal}{Phys. Rev. B}}
  \textbf{\bibinfo{volume}{102}}, \bibinfo{pages}{174504}
  (\bibinfo{year}{2020}).

\bibitem{Kezilebieke2020}
\bibinfo{author}{Kezilebieke, S.} \emph{et~al.}
\newblock \bibinfo{title}{Topological superconductivity in a designer
  ferromagnet-superconductor van der {Waals} heterostructure}.
\newblock \emph{\bibinfo{journal}{arXiv:2002.02141}}
  (\bibinfo{year}{2020}).

\bibitem{Yu1965}
\bibinfo{author}{Yu, L.}
\newblock \bibinfo{title}{Bound state in superconductors with paramagnetic
  impurities}.
\newblock \emph{\bibinfo{journal}{Acta Phys. Sin}}
  \textbf{\bibinfo{volume}{21}}, \bibinfo{pages}{75--91}
  (\bibinfo{year}{1965}).

\bibitem{Shiba1968}
\bibinfo{author}{Shiba, H.}
\newblock \bibinfo{title}{Classical spins in superconductors}.
\newblock \emph{\bibinfo{journal}{Prog. Theor. Phys.}}
  \textbf{\bibinfo{volume}{40}}, \bibinfo{pages}{435--451}
  (\bibinfo{year}{1968}).

\bibitem{Rusinov1969}
\bibinfo{author}{Rusinov, A.~I.}
\newblock \bibinfo{title}{Superconductivity near a paramagnetic impurity}.
\newblock \emph{\bibinfo{journal}{JETP Lett.}} \textbf{\bibinfo{volume}{9}},
  \bibinfo{pages}{85} (\bibinfo{year}{1969}).

\bibitem{Yazdani1997}
\bibinfo{author}{Yazdani, A.}, \bibinfo{author}{Jones, B.~A.},
  \bibinfo{author}{Lutz, C.~P.}, \bibinfo{author}{Crommie, M.~F.} \&
  \bibinfo{author}{Eigler, D.~M.}
\newblock \bibinfo{title}{Probing the local effects of magnetic impurities on
  superconductivity}.
\newblock \emph{\bibinfo{journal}{Science}} \textbf{\bibinfo{volume}{275}},
  \bibinfo{pages}{1767--1770} (\bibinfo{year}{1997}).

\bibitem{Ji2008}
\bibinfo{author}{Ji, S.~H.} \emph{et~al.}
\newblock \bibinfo{title}{High-resolution scanning tunneling spectroscopy of
  magnetic impurity induced bound states in the superconducting gap of {Pb}
  thin films}.
\newblock \emph{\bibinfo{journal}{Phys. Rev. Lett.}}
  \textbf{\bibinfo{volume}{100}}, \bibinfo{pages}{226801} (\bibinfo{year}{2008}).

\bibitem{Franke2011}
\bibinfo{author}{Franke, K.~J.}, \bibinfo{author}{Schulze, G.} \&
  \bibinfo{author}{Pascual, J.~I.}
\newblock \bibinfo{title}{Competition of superconducting phenomena and {Kondo}
  screening at the nanoscale}.
\newblock \emph{\bibinfo{journal}{Science}} \textbf{\bibinfo{volume}{332}},
  \bibinfo{pages}{940--944} (\bibinfo{year}{2011}).

\bibitem{Hatter2017}
\bibinfo{author}{Hatter, N.}, \bibinfo{author}{Heinrich, B.~W.},
  \bibinfo{author}{Rolf, D.} \& \bibinfo{author}{Franke, K.~J.}
\newblock \bibinfo{title}{Scaling of {Yu-Shiba-Rusinov} energies in the
  weak-coupling {Kondo} regime}.
\newblock \emph{\bibinfo{journal}{Nat. Commun.}} \textbf{\bibinfo{volume}{8}},
  \bibinfo{pages}{2016} (\bibinfo{year}{2017}).

\bibitem{Malavolti2018}
\bibinfo{author}{Malavolti, L.} \emph{et~al.}
\newblock \bibinfo{title}{Tunable spin--superconductor coupling of spin 1/2
  vanadyl phthalocyanine molecules}.
\newblock \emph{\bibinfo{journal}{Nano Lett.}} \textbf{\bibinfo{volume}{18}},
  \bibinfo{pages}{7955--7961} (\bibinfo{year}{2018}).

\bibitem{Senkpiel2019}
\bibinfo{author}{Senkpiel, J.} \emph{et~al.}
\newblock \bibinfo{title}{Robustness of {Yu-Shiba-Rusinov} resonances in the
  presence of a complex superconducting order parameter}.
\newblock \emph{\bibinfo{journal}{Phys. Rev. B}}
  \textbf{\bibinfo{volume}{100}}, \bibinfo{pages}{014502}
  (\bibinfo{year}{2019}).

\bibitem{Huang2019magnetic}
\bibinfo{author}{Huang, H.} \emph{et~al.}
\newblock \bibinfo{title}{Quantum phase transitions and the role of impurity-substrate hybridization in Yu-Shiba-Rusinov states}.
\newblock \emph{\bibinfo{journal}{Communications Physics}}
\textbf{\bibinfo{volume}{3}}, \bibinfo{pages}{199}
  (\bibinfo{year}{2020}).

\bibitem{Saldana2018}
\bibinfo{author}{Saldana, J.~C.} \emph{et~al.}
\newblock \bibinfo{title}{Two-impurity {Yu-Shiba-Rusinov} states in coupled
  quantum dots}.
\newblock \emph{\bibinfo{journal}{Phys. Rev. B}}
\textbf{\bibinfo{volume}{102}}, \bibinfo{pages}{195143}
  (\bibinfo{year}{2020}).

\bibitem{kim2018toward}
\bibinfo{author}{Kim, H.} \emph{et~al.}
\newblock \bibinfo{title}{Toward tailoring {Majorana} bound states in
  artificially constructed magnetic atom chains on elemental superconductors}.
\newblock \emph{\bibinfo{journal}{Sci. Adv.}} \textbf{\bibinfo{volume}{4}},
  \bibinfo{pages}{eaar5251} (\bibinfo{year}{2018}).

\bibitem{kamlapure2018engineering}
\bibinfo{author}{Kamlapure, A.}, \bibinfo{author}{Cornils, L.},
  \bibinfo{author}{Wiebe, J.} \& \bibinfo{author}{Wiesendanger, R.}
\newblock \bibinfo{title}{Engineering the spin couplings in atomically crafted
  spin chains on an elemental superconductor}.
\newblock \emph{\bibinfo{journal}{Nat. Commun.}} \textbf{\bibinfo{volume}{9}},
  \bibinfo{pages}{3253} (\bibinfo{year}{2018}).

\bibitem{Kitaev2001}
\bibinfo{author}{Kitaev, A.~Y.}
\newblock \bibinfo{title}{Unpaired {Majorana} fermions in quantum wires}.
\newblock \emph{\bibinfo{journal}{Physics-Uspekhi}}
  \textbf{\bibinfo{volume}{44}}, \bibinfo{pages}{131} (\bibinfo{year}{2001}).

\bibitem{Alicea2012}
\bibinfo{author}{Alicea, J.}
\newblock \bibinfo{title}{New directions in the pursuit of {Majorana} fermions
  in solid state systems}.
\newblock \emph{\bibinfo{journal}{Rep. Prog. Phys.}}
  \textbf{\bibinfo{volume}{75}}, \bibinfo{pages}{076501}
  (\bibinfo{year}{2012}).

\bibitem{Kezilebieke2018}
\bibinfo{author}{Kezilebieke, S.}, \bibinfo{author}{Dvorak, M.},
  \bibinfo{author}{Ojanen, T.} \& \bibinfo{author}{Liljeroth, P.}
\newblock \bibinfo{title}{Coupled {Yu-Shiba-Rusinov} states in molecular dimers
  on {NbSe2}}.
\newblock \emph{\bibinfo{journal}{Nano Lett.}} \textbf{\bibinfo{volume}{18}},
  \bibinfo{pages}{2311--2315} (\bibinfo{year}{2018}).

\bibitem{Assig2013}
\bibinfo{author}{Assig, M.} \emph{et~al.}
\newblock \bibinfo{title}{A 10\,{mK} scanning tunneling microscope operating in
  ultra high vacuum and high magnetic fields}.
\newblock \emph{\bibinfo{journal}{Rev. Sci. Instrum.}}
  \textbf{\bibinfo{volume}{84}}, \bibinfo{pages}{033903}
  (\bibinfo{year}{2013}).

\bibitem{Huang2020tunneling}
\bibinfo{author}{Huang, H.} \emph{et~al.}
\newblock \bibinfo{title}{Tunnelling dynamics between superconducting bound
  states at the atomic limit}.
\newblock \emph{\bibinfo{journal}{Nat. Phys.}} \textbf{\bibinfo{volume}{16}}, \bibinfo{pages}{1227} (\bibinfo{year}{2020}).

\bibitem{supinf}
\bibinfo{note}{See supplementary information}.

\bibitem{Balatsky2006}
\bibinfo{author}{Balatsky, A.~V.}, \bibinfo{author}{Vekhter, I.} \&
  \bibinfo{author}{Zhu, J.~X.}
\newblock \bibinfo{title}{Impurity-induced states in conventional and
  unconventional superconductors}.
\newblock \emph{\bibinfo{journal}{Rev. Mod. Phys.}}
  \textbf{\bibinfo{volume}{78}}, \bibinfo{pages}{373--433}
  (\bibinfo{year}{2006}).

\bibitem{Cornils2017}
\bibinfo{author}{Cornils, L.} \emph{et~al.}
\newblock \bibinfo{title}{Spin-resolved spectroscopy of the {Yu-Shiba-Rusinov}
  states of individual atoms}.
\newblock \emph{\bibinfo{journal}{Phys. Rev. Lett.}}
  \textbf{\bibinfo{volume}{119}}, \bibinfo{pages}{197002}
  (\bibinfo{year}{2017}).

\bibitem{schneider_atomic-scale_2020}
\bibinfo{author}{Schneider, L.}, \bibinfo{author}{Beck, P.},
  \bibinfo{author}{Wiebe, J.} \& \bibinfo{author}{Wiesendanger, R.}
\newblock \bibinfo{title}{Atomic-scale spin-polarization maps using
  functionalized superconducting probes}.
\newblock \emph{\bibinfo{journal}{arXiv:2006.05770}}
  (\bibinfo{year}{2020}).

\bibitem{Villas2020}
\bibinfo{author}{Villas, A.} \emph{et~al.}
\newblock \bibinfo{title}{Interplay between {Yu-Shiba-Rusinov} states and
  multiple {Andreev} reflections}.
\newblock \emph{\bibinfo{journal}{Phys. Rev. B}}
  \textbf{\bibinfo{volume}{101}}, \bibinfo{pages}{235445}
  (\bibinfo{year}{2020}).

\bibitem{Ruby2015a}
\bibinfo{author}{Ruby, M.} \emph{et~al.}
\newblock \bibinfo{title}{Tunneling processes into localized subgap states in
  superconductors}.
\newblock \emph{\bibinfo{journal}{Phys. Rev. Lett.}}
  \textbf{\bibinfo{volume}{115}}, \bibinfo{pages}{087001}
  (\bibinfo{year}{2015}).

\bibitem{Loth2012}
\bibinfo{author}{Loth, S.}, \bibinfo{author}{Baumann, S.},
  \bibinfo{author}{Lutz, C.~P.}, \bibinfo{author}{Eigler, D.} \&
  \bibinfo{author}{Heinrich, A.~J.}
\newblock \bibinfo{title}{Bistability in atomic-scale antiferromagnets}.
\newblock \emph{\bibinfo{journal}{Science}} \textbf{\bibinfo{volume}{335}},
  \bibinfo{pages}{196--199} (\bibinfo{year}{2012}).

\bibitem{Ast2016}
\bibinfo{author}{Ast, C.~R.} \emph{et~al.}
\newblock \bibinfo{title}{Sensing the quantum limit in scanning tunnelling
  spectroscopy}.
\newblock \emph{\bibinfo{journal}{Nat. Commun.}} \textbf{\bibinfo{volume}{7}},
  \bibinfo{pages}{13009} (\bibinfo{year}{2016}).

\bibitem{Devoret1990}
\bibinfo{author}{Devoret, M.~H.} \emph{et~al.}
\newblock \bibinfo{title}{Effect of the electromagnetic environment on the
  {Coulomb} blockade in ultrasmall tunnel-junctions}.
\newblock \emph{\bibinfo{journal}{Phys. Rev. Lett.}}
  \textbf{\bibinfo{volume}{64}}, \bibinfo{pages}{1824--1827}
  (\bibinfo{year}{1990}).

\bibitem{Averin1990}
\bibinfo{author}{Averin, D.~V.}, \bibinfo{author}{Nazarov, Y.~V.} \&
  \bibinfo{author}{Odintsov, A.~A.}
\newblock \bibinfo{title}{Incoherent tunneling of the {Cooper} pairs and
  magnetic-flux quanta in ultrasmall {Josephson}-junctions}.
\newblock \emph{\bibinfo{journal}{Physica B}} \textbf{\bibinfo{volume}{165}},
  \bibinfo{pages}{945--946} (\bibinfo{year}{1990}).

\bibitem{Ingold1994}
\bibinfo{author}{Ingold, G.~L.}, \bibinfo{author}{Grabert, H.} \&
  \bibinfo{author}{Eberhardt, U.}
\newblock \bibinfo{title}{{Cooper}-pair current through ultrasmall
  {Josephson}-junctions}.
\newblock \emph{\bibinfo{journal}{Phys. Rev. B}} \textbf{\bibinfo{volume}{50}},
  \bibinfo{pages}{395--402} (\bibinfo{year}{1994}).

\bibitem{pientka2013}
\bibinfo{author}{Pientka, F.}, \bibinfo{author}{Glazman, L.~I.} \&
  \bibinfo{author}{von Oppen, F.}
\newblock \bibinfo{title}{Topological superconducting phase in helical {Shiba}
  chains}.
\newblock \emph{\bibinfo{journal}{Phys. Rev. B}} \textbf{\bibinfo{volume}{88}},
  \bibinfo{pages}{155420} (\bibinfo{year}{2013}).

\bibitem{Pientka2014}
\bibinfo{author}{Pientka, F.}, \bibinfo{author}{Glazman, L.~I.} \&
  \bibinfo{author}{von Oppen, F.}
\newblock \bibinfo{title}{Unconventional topological phase transitions in
  helical {Shiba} chains}.
\newblock \emph{\bibinfo{journal}{Phys. Rev. B}} \textbf{\bibinfo{volume}{89}},
  \bibinfo{pages}{180505} (\bibinfo{year}{2014}).

\bibitem{Hoffman2016}
\bibinfo{author}{Hoffman, S.}, \bibinfo{author}{Klinovaja, J.} \&
  \bibinfo{author}{Loss, D.}
\newblock \bibinfo{title}{Topological phases of inhomogeneous
  superconductivity}.
\newblock \emph{\bibinfo{journal}{Phys. Rev. B}} \textbf{\bibinfo{volume}{93}},
  \bibinfo{pages}{165418} (\bibinfo{year}{2016}).

\bibitem{Andolina2017}
\bibinfo{author}{Andolina, G.~M.} \& \bibinfo{author}{Simon, P.}
\newblock \bibinfo{title}{Topological properties of chains of magnetic
  impurities on a superconducting substrate: Interplay between the {Shiba} band
  and ferromagnetic wire limits}.
\newblock \emph{\bibinfo{journal}{Phys. Rev. B}} \textbf{\bibinfo{volume}{96}},
  \bibinfo{pages}{235411} (\bibinfo{year}{2017}).

\bibitem{Klinovaja2013}
\bibinfo{author}{Klinovaja, J.}, \bibinfo{author}{Stano, P.},
  \bibinfo{author}{Yazdani, A.} \& \bibinfo{author}{Loss, D.}
\newblock \bibinfo{title}{Topological superconductivity and {Majorana} fermions
  in {RKKY} systems}.
\newblock \emph{\bibinfo{journal}{Phys. Rev. Lett.}}
  \textbf{\bibinfo{volume}{111}}, \bibinfo{pages}{186805}
  (\bibinfo{year}{2013}).

\bibitem{vazifeh2013self}
\bibinfo{author}{Vazifeh, M.} \& \bibinfo{author}{Franz, M.}
\newblock \bibinfo{title}{Self-organized topological state with {Majorana}
  fermions}.
\newblock \emph{\bibinfo{journal}{Phys. Rev. Lett.}}
  \textbf{\bibinfo{volume}{111}}, \bibinfo{pages}{206802}
  (\bibinfo{year}{2013}).

\bibitem{Braunecker2013}
\bibinfo{author}{Braunecker, B.} \& \bibinfo{author}{Simon, P.}
\newblock \bibinfo{title}{Interplay between classical magnetic moments and
  superconductivity in quantum one-dimensional conductors: Toward a
  self-sustained topological majorana phase}.
\newblock \emph{\bibinfo{journal}{Phys. Rev. Lett.}}
  \textbf{\bibinfo{volume}{111}}, \bibinfo{pages}{147202}
  (\bibinfo{year}{2013}).

\bibitem{eschrig_theory_2003}
\bibinfo{author}{Eschrig, M.}, \bibinfo{author}{Kopu, J.},
  \bibinfo{author}{Cuevas, J.~C.} \& \bibinfo{author}{Schön, G.}
\newblock \bibinfo{title}{Theory of half-metal/superconductor
  heterostructures}.
\newblock \emph{\bibinfo{journal}{Phys. Rev. Lett.}}
  \textbf{\bibinfo{volume}{90}}, \bibinfo{pages}{137003}
  (\bibinfo{year}{2003}).

\end{thebibliography}

\begin{thebibliography}{10}
\expandafter\ifx\csname url\endcsname\relax
  \def\url#1{\texttt{#1}}\fi
\expandafter\ifx\csname urlprefix\endcsname\relax\def\urlprefix{URL }\fi
\providecommand{\bibinfo}[2]{#2}
\providecommand{\eprint}[2][]{\url{#2}}

\bibitem{si_Huang2020tunneling}
\bibinfo{author}{Huang, H.} \emph{et~al.}
\newblock \bibinfo{title}{Tunnelling dynamics between superconducting bound
  states at the atomic limit}.
\newblock \emph{\bibinfo{journal}{Nat. Phys.}} \textbf{\bibinfo{volume}{16}},
  \bibinfo{pages}{1227}  (\bibinfo{year}{2020}).

\bibitem{si_Sekula1972}
\bibinfo{author}{Sekula, S.} \& \bibinfo{author}{Kernohan, R.}
\newblock \bibinfo{title}{Magnetic properties of superconducting vanadium}.
\newblock \emph{\bibinfo{journal}{Phys. Rev. B}} \textbf{\bibinfo{volume}{5}},
  \bibinfo{pages}{904} (\bibinfo{year}{1972}).

\bibitem{si_Zasadzinski1982}
\bibinfo{author}{Zasadzinski, J.}, \bibinfo{author}{Burnell, D.},
  \bibinfo{author}{Wolf, E.} \& \bibinfo{author}{Arnold, G.}
\newblock \bibinfo{title}{Superconducting tunneling study of vanadium}.
\newblock \emph{\bibinfo{journal}{Phys. Rev. B}} \textbf{\bibinfo{volume}{25}},
  \bibinfo{pages}{1622} (\bibinfo{year}{1982}).

\bibitem{si_Jensen1982}
\bibinfo{author}{Jensen, V.}, \bibinfo{author}{Andersen, J.~N.},
  \bibinfo{author}{Nielsen, H.~B.} \& \bibinfo{author}{Adams, D.~L.}
\newblock \bibinfo{title}{The surface-structure of {V(100)}}.
\newblock \emph{\bibinfo{journal}{Surf. Sci.}} \textbf{\bibinfo{volume}{116}},
  \bibinfo{pages}{66--84} (\bibinfo{year}{1982}).

\bibitem{si_Koller2001}
\bibinfo{author}{Koller, R.} \emph{et~al.}
\newblock \bibinfo{title}{The structure of the oxygen induced (1$\times$5)
  reconstruction of {V}(100)}.
\newblock \emph{\bibinfo{journal}{Surf. Sci.}} \textbf{\bibinfo{volume}{480}},
  \bibinfo{pages}{11--24} (\bibinfo{year}{2001}).

\bibitem{si_Dulot2001}
\bibinfo{author}{Dulot, F.} \emph{et~al.}
\newblock \bibinfo{title}{{V}(001) surface structures analysed by {RHEED} and
  {STM}}.
\newblock \emph{\bibinfo{journal}{Surf. Sci.}} \textbf{\bibinfo{volume}{473}},
  \bibinfo{pages}{172--182} (\bibinfo{year}{2001}).

\bibitem{si_Jack2016}
\bibinfo{author}{Jack, B.} \emph{et~al.}
\newblock \bibinfo{title}{Critical {Josephson} current in the dynamical
  {Coulomb} blockade regime}.
\newblock \emph{\bibinfo{journal}{Phys. Rev. B}} \textbf{\bibinfo{volume}{93}},
  \bibinfo{pages}{020504} (\bibinfo{year}{2016}).

\bibitem{si_Cornils2017}
\bibinfo{author}{Cornils, L.} \emph{et~al.}
\newblock \bibinfo{title}{Spin-resolved spectroscopy of the {Yu-Shiba-Rusinov}
  states of individual atoms}.
\newblock \emph{\bibinfo{journal}{Phys. Rev. Lett.}}
  \textbf{\bibinfo{volume}{119}}, \bibinfo{pages}{197002}
  (\bibinfo{year}{2017}).

\bibitem{si_Villas2020}
\bibinfo{author}{Villas, A.} \emph{et~al.}
\newblock \bibinfo{title}{Interplay between {Yu-Shiba-Rusinov} states and
  multiple {Andreev} reflections}.
\newblock \emph{\bibinfo{journal}{Phys. Rev. B}}
  \textbf{\bibinfo{volume}{101}}, \bibinfo{pages}{235445}
  (\bibinfo{year}{2020}).

\bibitem{si_Ruby2015a}
\bibinfo{author}{Ruby, M.} \emph{et~al.}
\newblock \bibinfo{title}{Tunneling processes into localized subgap states in
  superconductors}.
\newblock \emph{\bibinfo{journal}{Phys. Rev. Lett.}}
  \textbf{\bibinfo{volume}{115}}, \bibinfo{pages}{087001}
  (\bibinfo{year}{2015}).

\bibitem{si_randeria_scanning_2016}
\bibinfo{author}{Randeria, M.~T.}, \bibinfo{author}{Feldman, B.~E.},
  \bibinfo{author}{Drozdov, I.~K.} \& \bibinfo{author}{Yazdani, A.}
\newblock \bibinfo{title}{Scanning {Josephson} spectroscopy on the atomic
  scale}.
\newblock \emph{\bibinfo{journal}{Phys. Rev. B}} \textbf{\bibinfo{volume}{93}},
  \bibinfo{pages}{161115} (\bibinfo{year}{2016}).

\bibitem{si_Shiba1968}
\bibinfo{author}{Shiba, H.}
\newblock \bibinfo{title}{Classical spins in superconductors}.
\newblock \emph{\bibinfo{journal}{Prog. Theor. Phys.}}
  \textbf{\bibinfo{volume}{40}}, \bibinfo{pages}{435--451}
  (\bibinfo{year}{1968}).

\bibitem{si_Balatsky2006}
\bibinfo{author}{Balatsky, A.~V.}, \bibinfo{author}{Vekhter, I.} \&
  \bibinfo{author}{Zhu, J.~X.}
\newblock \bibinfo{title}{Impurity-induced states in conventional and
  unconventional superconductors}.
\newblock \emph{\bibinfo{journal}{Rev. Mod. Phys.}}
  \textbf{\bibinfo{volume}{78}}, \bibinfo{pages}{373--433}
  (\bibinfo{year}{2006}).

\bibitem{si_Ast2016}
\bibinfo{author}{Ast, C.~R.} \emph{et~al.}
\newblock \bibinfo{title}{Sensing the quantum limit in scanning tunnelling
  spectroscopy}.
\newblock \emph{\bibinfo{journal}{Nat. Commun.}} \textbf{\bibinfo{volume}{7}},
  \bibinfo{pages}{13009} (\bibinfo{year}{2016}).

\bibitem{si_Devoret1990}
\bibinfo{author}{Devoret, M.~H.} \emph{et~al.}
\newblock \bibinfo{title}{Effect of the electromagnetic environment on the
  {Coulomb} blockade in ultrasmall tunnel-junctions}.
\newblock \emph{\bibinfo{journal}{Phys. Rev. Lett.}}
  \textbf{\bibinfo{volume}{64}}, \bibinfo{pages}{1824--1827}
  (\bibinfo{year}{1990}).

\bibitem{si_Averin1990}
\bibinfo{author}{Averin, D.~V.}, \bibinfo{author}{Nazarov, Y.~V.} \&
  \bibinfo{author}{Odintsov, A.~A.}
\newblock \bibinfo{title}{Incoherent tunneling of the {Cooper} pairs and
  magnetic-flux quanta in ultrasmall {Josephson}-junctions}.
\newblock \emph{\bibinfo{journal}{Physica B}} \textbf{\bibinfo{volume}{165}},
  \bibinfo{pages}{945--946} (\bibinfo{year}{1990}).

\bibitem{si_Ingold1994}
\bibinfo{author}{Ingold, G.~L.}, \bibinfo{author}{Grabert, H.} \&
  \bibinfo{author}{Eberhardt, U.}
\newblock \bibinfo{title}{{Cooper}-pair current through ultrasmall
  {Josephson}-junctions}.
\newblock \emph{\bibinfo{journal}{Phys. Rev. B}} \textbf{\bibinfo{volume}{50}},
  \bibinfo{pages}{395--402} (\bibinfo{year}{1994}).

\bibitem{si_Cuevas1996}
\bibinfo{author}{Cuevas, J.~C.}, \bibinfo{author}{Mart\'{\i}n-Rodero, A.} \&
  \bibinfo{author}{Yeyati, A.~L.}
\newblock \bibinfo{title}{Hamiltonian approach to the transport properties of
  superconducting quantum point contacts}.
\newblock \emph{\bibinfo{journal}{Phys. Rev. B}} \textbf{\bibinfo{volume}{54}},
  \bibinfo{pages}{7366--7379} (\bibinfo{year}{1996}).

\bibitem{si_Zhu2016}
\bibinfo{author}{Zhu, J.-X.}
\newblock \emph{\bibinfo{title}{Bogoliubov-de Gennes method and its
  applications}}, vol. \bibinfo{volume}{924} (\bibinfo{publisher}{Springer},
  \bibinfo{year}{2016}).

\bibitem{si_Nadj2013}
\bibinfo{author}{Nadj-Perge, S.}, \bibinfo{author}{Drozdov, I.~K.},
  \bibinfo{author}{Bernevig, B.~A.} \& \bibinfo{author}{Yazdani, A.}
\newblock \bibinfo{title}{Proposal for realizing {Majorana} fermions in chains
  of magnetic atoms on a superconductor}.
\newblock \emph{\bibinfo{journal}{Phys. Rev. B}} \textbf{\bibinfo{volume}{88}},
  \bibinfo{pages}{020407} (\bibinfo{year}{2013}).

\bibitem{si_Pientka2013}
\bibinfo{author}{Pientka, F.}, \bibinfo{author}{Glazman, L.~I.} \&
  \bibinfo{author}{von Oppen, F.}
\newblock \bibinfo{title}{Topological superconducting phase in helical {Shiba}
  chains}.
\newblock \emph{\bibinfo{journal}{Phys. Rev. B}} \textbf{\bibinfo{volume}{88}},
  \bibinfo{pages}{155420} (\bibinfo{year}{2013}).

\bibitem{si_Pientka2014}
\bibinfo{author}{Pientka, F.}, \bibinfo{author}{Glazman, L.~I.} \&
  \bibinfo{author}{von Oppen, F.}
\newblock \bibinfo{title}{Unconventional topological phase transitions in
  helical {Shiba} chains}.
\newblock \emph{\bibinfo{journal}{Phys. Rev. B}} \textbf{\bibinfo{volume}{89}},
  \bibinfo{pages}{180505} (\bibinfo{year}{2014}).

\bibitem{si_Li2014}
\bibinfo{author}{Li, J.} \emph{et~al.}
\newblock \bibinfo{title}{Topological superconductivity induced by
  ferromagnetic metal chains}.
\newblock \emph{\bibinfo{journal}{Phys. Rev. B}} \textbf{\bibinfo{volume}{90}},
  \bibinfo{pages}{235433} (\bibinfo{year}{2014}).

\bibitem{si_Heimes2014}
\bibinfo{author}{Heimes, A.}, \bibinfo{author}{Kotetes, P.} \&
  \bibinfo{author}{Sch\"on, G.}
\newblock \bibinfo{title}{Majorana fermions from {Shiba} states in an
  antiferromagnetic chain on top of a superconductor}.
\newblock \emph{\bibinfo{journal}{Phys. Rev. B}} \textbf{\bibinfo{volume}{90}},
  \bibinfo{pages}{060507} (\bibinfo{year}{2014}).

\bibitem{si_Hoffman2016}
\bibinfo{author}{Hoffman, S.}, \bibinfo{author}{Klinovaja, J.} \&
  \bibinfo{author}{Loss, D.}
\newblock \bibinfo{title}{Topological phases of inhomogeneous
  superconductivity}.
\newblock \emph{\bibinfo{journal}{Phys. Rev. B}} \textbf{\bibinfo{volume}{93}},
  \bibinfo{pages}{165418} (\bibinfo{year}{2016}).

\bibitem{si_Poyhonen2016}
\bibinfo{author}{P\"oyh\"onen, K.}, \bibinfo{author}{Weststr\"om, A.} \&
  \bibinfo{author}{Ojanen, T.}
\newblock \bibinfo{title}{Topological superconductivity in ferromagnetic atom
  chains beyond the deep-impurity regime}.
\newblock \emph{\bibinfo{journal}{Phys. Rev. B}} \textbf{\bibinfo{volume}{93}},
  \bibinfo{pages}{014517} (\bibinfo{year}{2016}).

\bibitem{si_Andolina2017}
\bibinfo{author}{Andolina, G.~M.} \& \bibinfo{author}{Simon, P.}
\newblock \bibinfo{title}{Topological properties of chains of magnetic
  impurities on a superconducting substrate: Interplay between the {Shiba} band
  and ferromagnetic wire limits}.
\newblock \emph{\bibinfo{journal}{Phys. Rev. B}} \textbf{\bibinfo{volume}{96}},
  \bibinfo{pages}{235411} (\bibinfo{year}{2017}).

\bibitem{si_Theiler2019}
\bibinfo{author}{Theiler, A.}, \bibinfo{author}{Bj\"ornson, K.} \&
  \bibinfo{author}{Black-Schaffer, A.~M.}
\newblock \bibinfo{title}{Majorana bound state localization and energy
  oscillations for magnetic impurity chains on conventional superconductors}.
\newblock \emph{\bibinfo{journal}{Phys. Rev. B}}
  \textbf{\bibinfo{volume}{100}}, \bibinfo{pages}{214504}
  (\bibinfo{year}{2019}).

\bibitem{si_Nadj-Perge2014}
\bibinfo{author}{Nadj-Perge, S.} \emph{et~al.}
\newblock \bibinfo{title}{Observation of {Majorana} fermions in ferromagnetic
  atomic chains on a superconductor}.
\newblock \emph{\bibinfo{journal}{Science}} \textbf{\bibinfo{volume}{346}},
  \bibinfo{pages}{602--607} (\bibinfo{year}{2014}).

\bibitem{si_Klinovaja2013}
\bibinfo{author}{Klinovaja, J.}, \bibinfo{author}{Stano, P.},
  \bibinfo{author}{Yazdani, A.} \& \bibinfo{author}{Loss, D.}
\newblock \bibinfo{title}{Topological superconductivity and {Majorana} fermions
  in {RKKY} systems}.
\newblock \emph{\bibinfo{journal}{Phys. Rev. Lett.}}
  \textbf{\bibinfo{volume}{111}}, \bibinfo{pages}{186805}
  (\bibinfo{year}{2013}).

\bibitem{si_Vazifeh2013}
\bibinfo{author}{Vazifeh, M.~M.} \& \bibinfo{author}{Franz, M.}
\newblock \bibinfo{title}{Self-organized topological state with {Majorana}
  fermions}.
\newblock \emph{\bibinfo{journal}{Phys. Rev. Lett.}}
  \textbf{\bibinfo{volume}{111}}, \bibinfo{pages}{206802}
  (\bibinfo{year}{2013}).

\bibitem{si_Braunecker2013}
\bibinfo{author}{Braunecker, B.} \& \bibinfo{author}{Simon, P.}
\newblock \bibinfo{title}{Interplay between classical magnetic moments and
  superconductivity in quantum one-dimensional conductors: Toward a
  self-sustained topological {Majorana} phase}.
\newblock \emph{\bibinfo{journal}{Phys. Rev. Lett.}}
  \textbf{\bibinfo{volume}{111}}, \bibinfo{pages}{147202}
  (\bibinfo{year}{2013}).

\bibitem{si_Jack2019}
\bibinfo{author}{J{\"a}ck, B.} \emph{et~al.}
\newblock \bibinfo{title}{Observation of a {Majorana} zero mode in a
  topologically protected edge channel}.
\newblock \emph{\bibinfo{journal}{Science}} \textbf{\bibinfo{volume}{364}},
  \bibinfo{pages}{1255--1259} (\bibinfo{year}{2019}).

\end{thebibliography}
\end{document}